\documentclass[amsmath,amssymb,amsfonts, superscriptaddress, aps, twocolumn]{revtex4-1}
\usepackage[german,american,english]{babel}
\usepackage{graphicx}
\usepackage{graphics}
\usepackage{dcolumn}
\usepackage{bm}
\usepackage{amssymb}
\usepackage{amsmath}
\usepackage{amsfonts}
\usepackage{epsfig}
\usepackage{color}
\usepackage{ulem}
\usepackage{hyperref}
\hypersetup{
    colorlinks,
    citecolor=blue,
    filecolor=black,
    linkcolor=blue,
    urlcolor=blue
}

\newcommand {\bra} [1] {\langle #1 |}
\newcommand {\ket} [1] {| #1 \rangle}
\newcommand {\bkt} [1] {\langle #1 \rangle}
\newcommand {\dbkt} [2] {\langle #1 | #2 \rangle}
\newcommand {\tbkt} [3] {\langle #1 | #2 | #3 \rangle}
\newcommand {\pd} [2] {\frac{\partial #1}{\partial #2}}
\newcommand {\td} [2] {\frac{d #1}{d #2}}

 \newcommand {\beq}{\begin{equation}}
\newcommand {\eeq}{\end{equation}}

\begin{document}
\title{Inter-band Coherence Response to Electric Fields in Crystals:\\ Berry-Phase Contributions and Disorder Effects}

\author{Dimitrie Culcer}
\affiliation{School of Physics and Australian Research Council Centre of Excellence in Low-Energy Electronics Technologies, UNSW Node, The University of New South Wales, Sydney 2052, Australia}

\author{Akihiko Sekine}
\affiliation{Department of Physics, The University of Texas at Austin, Austin, Texas 78712, USA}

\author{Allan H. MacDonald}
\affiliation{Department of Physics, The University of Texas at Austin, Austin, Texas 78712, USA}

\begin{abstract}
In solid state conductors, linear response to a steady electric field is normally dominated by Bloch state occupation number changes that are correlated with group velocity and lead to a steady state current. Recently it has been realized that, for a number of important physical observables, the most important response even in conductors can be electric-field induced coherence between Bloch states in different bands, such as that responsible for screening in dielectrics. Examples include the anomalous and spin-Hall effects, spin torques in magnetic conductors, and
the minimum conductivity and chiral anomaly in Weyl and Dirac semimetals. 
In this paper we present a general quantum kinetic theory of linear response to an electric field which can be applied to solids with arbitrarily complicated band structures and includes the inter-band coherence response and the Bloch-state repopulation responses on an equal footing. One of the principal aims of our work is to enable extensive transport theory applications using computational packages constructed in terms of maximally localized Wannier functions. 
To this end we provide a complete correspondence between the Bloch and Wannier formulations of our theory. 
The formalism is based on density-matrix equations of motion, on a Born approximation treatment of disorder, and on 
an expansion in scattering rate to leading non-trivial order.  Our use of a Born approximation omits some 
physical effects and represents a compromise between comprehensiveness and practicality.  
The quasiparticle bands are treated in a completely general manner that allows for arbitrary 
forms of the spin-orbit interaction and for the broken time reversal symmetry of magnetic conductors. 
We demonstrate that the inter-band response in conductors consists primarily of two terms: an intrinsic contribution due 
to the entire Fermi sea that captures, among other effects, the Berry curvature contribution to wave-packet 
dynamics, and an anomalous contribution caused by scattering that is sensitive to the presence of the Fermi surface. 
To demonstrate the rich physics captured by our theory, 
we explicitly solve for some electric-field response properties of simple model systems
that are known to be dominated by interband coherence contributions. 
At the same time we discuss an extensive list of complicated problems that cannot be solved analytically. 
Our goal is to stimulate progress in computational transport theory for electrons in crystals.  
\end{abstract}

\maketitle


\section{Introduction}
In response to a steady electric field $\bm{E}$, weakly disordered Fermi-liquid conductors reach a steady state that does not 
break translational symmetry. After disorder averaging, such a steady state can be completely characterized by a non-equilibrium 
single-particle density-matrix, $\rho_{n',n}(\bm{k})$ diagonal in the Bloch state wave-vector $\bm{k}$ but not 
in the equilibrium band index $n$. In general all single-particle observables $O$ maintain their 
crystal periodicity when they respond to $\bm{E}$ and therefore have expectation values of the form  
\begin{equation} 
\langle O \rangle = \sum_{\bm{k},n,n'} \, \langle n,\bm{k}|O|n',\bm{k}\rangle \, \rho_{n',n}(\bm{k}).  
\end{equation}
In metals linear response to a constant, uniform electric field is normally
dominated by the band-diagonal part of the density matrix, namely $n' = n$, which represents a change in the 
occupation probability of the Bloch states. 
The past two decades have nevertheless
 provided many examples of observables in conductors whose linear response to a steady electric field is dominated by the inter-band coherence ($n' \ne n$) response that is normally important only in dielectrics. Examples of linear response quantities that are often in this category include, but are not limited to, the quasiparticle spin-density 
 response responsible for spin-orbit and spin-transfer torques in ferromagnets \cite{StilesZangwill_PRB02, TataraKohno_PRL04, Tserkovnyak_MgnRMP05, Ralph2008, Haney_JMMM08, RalphStiles, Manchon_SOTorque_PRB09, Culcer_Fmg_SpinTransfer_PRB09, Garate_TI/FM_InverseSpinGalv_PRL10, Miron_SOTorque_NM10, Brataas_SOTorque_NN14, Mellnik_Nature14}, the anomalous Hall effect in magnetic conductors \cite{Luttinger_AHE_PR58, Smit_SS_58, Berger_SJ_PRB1970, NozLew_SSSJ_JP73, Dugaev_AHE_FM_Localization_PRB01, Crepieux_AHE_KuboDirac_PRB01, Jungwirth_PRL02, Culcer_AHE_PRB03, AHE_vertex_PRL_2006, Sinitsyn_AHE_KuboStreda_PRB07, Nagaosa_AHE_PRB08, Sinitsyn_AHE_Review_JPCM08, Kovalev_Multiband_AHE_PRB09, Nagaosa-AHE-2010, Yang_SJ_SctUniversality_PRB10, Tse_TI_MOKE_PRL2010, ZangNagaosa_TI_Monopole_PRB10, Nomura_PRL_2011, Culcer_TI_AHE_PRB11, Chang_TIF_ferromagnetism_AHE_AM2013}, the spin-Hall effect in paramagnetic semiconductors and metals \cite{Hirsch_SHE_99, Schwab_SHE_EPJB02, Murakami_SHE_Sci03, Sinova2004, Inoue_RashbaSHE_Vertex_PRB04, Murakami_SHI_PRL04, Dimitrova_05, Chalaev_SHE_PRB05, Shytov_SHE_PRB06, Khateskii_NoSHE_PRL06, CulcerWinkler_NonSHE_PRL07, Culcer_SteadyState_PRB07, Sinova2015}, the minimum conductivity of graphene, graphene multilayers \cite{Novoselov_Science04, Novoselov_Nature05, Geim_NM07, Katsnelson_Zbw_MinCond_EPJB06, Adam_Gfn_PNAS07, Chen_NP08, Zhang_NP09, PhysRevB.76.245405, CastroNeto2009, PhysRevB.87.035415, PhysRevLett.99.246803, PhysRevLett.100.046403, Mishchenko_EPL08, Culcer_Gfn_Transp_PRB08, Culcer_Bil_PRB09, Trushin_BLG_MinCond_PRB10, SDS_Gfn_RMP11} and topological insulator surface states \cite{Culcer_TI_Kineq_PRB10, Kim_TI_Gate_MinCond_NP12, Adam_TI_Tnsp_PRB12, Qiuzi_TI_Tnsp_PRB12, Tkachov_TI_Review_PSS13}, the chiral anomalies of Weyl and Dirac semimetals \cite{PhysRevLett.113.246402, PhysRevX.5.031023, Shekhar_NP15, Xiong_Science15, Li2015, Liang_NM16, Burkov2015, Li_NC16, Hellerstedt_NL16, Lu_FP16, Wang_NC16}
and interaction effects \cite{Culcer_TI_Int_PRB11, Weizhe_TITF_2014_prb} including Coulomb drag, in particular in chiral materials \cite{RevModPhys.88.025003, Zheng-2DES-1993, edge_drag_prb, Tse2007, Polini_graphene_screening_SSC2007, Badalyan_SpinHallDrag, Hwang2011, M.Carrega2012, Nandi_eh_nat, Kim2012, Gorbachevi_nature_2012, 1D-1D_science_Luttinger, Drag_graphnen_nl, Hong_drag_2015}. 
Related examples may be found in ultracold atomic gases.
In this paper we refer to band off-diagonal density-matrix response to a steady electric field generically as anomalous response.

When calculating the linear response of these observables to an electric field, accounting for the interplay between diagonal and off-diagonal density matrix response is vital in order to capture the underlying physics and determine the correct result. The response properties in which we are interested typically involve a competition between intrinsic inter-band coherence effects dependent only on the band structure of the crystal, and intra-band response that is limited by scattering by electron-phonon or electron-magnon interactions or by crystal imperfections. The interplay
has most often been addressed theoretically using simplified models with narrow applicability, or by using relaxation time approximations that typically fail to account accurately for inter-band coherence and complex Fermi surface topologies. 
Yet many systems of interest, such as metals and a host of recently discovered topological materials, have complicated band structures for which few-band models are simply unavailable or unsuitable,
and accurate solutions can only be found only using computational techniques. 
Frequently even the Berry curvatures can only be calculated numerically, and incorporating inter-band coherence effects due to disorder is even more challenging. 
Direct numerical implementation of transport calculations is especially challenging in the DC limit where the frequency characterizing the time-dependence of the external electric field tends towards zero.

Motivated by these observations, the goal of this paper, which generalizes Refs.~\onlinecite{Vasko, Zubarev}, is to devise a general transport formalism of broad applicability that accounts for the diagonal and off-diagonal responses, both intrinsic and extrinsic, on an equal footing. The results we present are suitable for use in numerical calculations that allow for general Fermi surface topologies involving many pockets with irregular shapes. In this way our work generalizes to the case of anomalous response transport theories that account for Fermi surface peculiarities, and for realistic scattering properties on those Fermi surfaces that are not accurately captured by simplifying relaxation time approximations. Because it requires efficient evaluation of velocity-operator matrix elements in a representation of orthonormal Bloch states, our intention is that applications of our transport formalism take advantage of recent progress in advancing maximally-localised Wannier orbital tools \cite{Wannier-1,Wannier-2} for constructing accurate representations of crystal Hamiltonians. To this end we provide in the present work two distinct but related sets of expressions: one set is formulated in the crystal momentum representation, which has been the natural language of conventional transport theory, while an equivalent set is expressed in the Wannier representation, which is the natural language of tight-binding models. 

Theories of the transport steady state must account for whichever Bloch state scattering mechanisms play the dominant role in limiting the repopulation of states near the Fermi level. The theory we present treats weak elastic disorder in the Born approximation and assumes that the Wannier representation of the crystal's $\bm{k} \cdot \bm{p}$ Hamiltonian is known. We focus on the off-diagonal response of the Bloch state density matrix which is normally dropped in theories of electronic transport in metals. Whereas our theory is formulated in the spirit of earlier works by Karplus, Luttinger and Kohn, we make full use of insights acquired in the last four decades, in particular in the identification of topological terms in linear response stemming from the Berry curvature of Bloch states. At the same time our response theory retains, on the same footing as the Berry-curvature terms, contributions to leading sub-dominant order in a weak scattering expansion of the density matrix. In this introductory paper we focus on the case of elastic scattering from crystal imperfections which we characterize by the variance of Wannier representation disorder matrix elements, or equivalently through an average over impurity configurations. Nevertheless the theory is presented in a form that can straightforwardly be extended to account for (i) time-dependent external fields and disorder potentials (ii) inelastic scattering (iii) more complex scattering mechanisms such as skew scattering and side jump that require going beyond the conventional Born approximation and (iv) more complex averages over impurity configurations including e.g. terms with crossing impurity lines describing interference between scattering at different sites \cite{Ado2016}. 

The kinetic equations have a complex matrix structure, in particular those involving the interband part of the density matrix. The band-diagonal response contains the usual Fermi-surface response which diverges when disorder vanishes. The off-diagonal response is driven by the intrinsic band structure, which includes Berry curvature contributions, and also indirectly by a scattering term involving only quasiparticles at the Fermi surface. These will be referred to as the \textit{off-diagonal driving term} and the \textit{anomalous driving term} respectively.

We note that the band off-diagonal response studied in this work represents inter-band coherence in the same way that off-diagonal terms in the density matrix of a quantum bit are associated with coherence between the bit \textit{up} and \textit{down} states. Likewise, the understanding of the dephasing time in quantum computation and solid state transport is conceptually similar, and in both cases is related to time-dependent perturbations. Nevertheless, the bit is invariably a localised system whereas the situations of interest in this paper concern extended systems having a Fermi surface, for which the steady state is qualitatively different from e.g. transport through a quantum dot or superconducting island. The notion of a dephasing time relevant to our work is the same as that encountered in weak localization. Whereas weak localization is a coherence effect induced by impurity scattering, in which one is interested in coherence between states in the same band but with different wave vectors, the relevant concept in the present work is coherence between states from different bands induced by an external electric field.

Our paper is organized as follows. In Section \ref{BW} we introduce the Hamiltonian and density matrix in the Bloch and Wannier representations. In Section \ref{Disorder} we discuss a variety of related models for disorder in a crystal, which provide a context for relating disorder scattering matrix elements to the wave-vector dependence of the 
orbital content of Bloch states.  In Section \ref{DM} we use the Born approximation to derive the form of the collision term in a general kinetic equation for the full density matrix. Finally in Section \ref{LRT} we present a theory for the response of the Bloch-state density-matrix of a crystal to a spatially constant electric field. The results obtained and their implications are discussed in Section \ref{Disc}. In Section \ref{Appl} we discuss some examples of material/observable combinations for which off-diagonal response is often important.   The paper concludes with a summary of our 
conclusions and an outlook for future work.  

\section{Bloch and Wannier Representations}
\label{BW}

Transport theory is most conveniently formulated in terms of momentum-space orbitals, whereas disorder potentials 
are best characterized in terms of their real-space orbital matrix elements.  Partly for this reason,
we use both Bloch and Wannier representations of crystalline wavefunctions throughout this paper.
The two-types of states are related by 
\begin{equation}\label{BlochWann}
\arraycolsep 0.3ex
\begin{array}{rl}
\displaystyle |i,\bm{k}\rangle = & \displaystyle \frac{1}{\sqrt{N}}  \sum_{\bm{L}}  \exp( i \bm{k} \cdot \bm{L})  \; |i,\bm{L}\rangle, \\\bm
\displaystyle |i,\bm{L}\rangle = & \displaystyle \frac{1}{\sqrt{N}}   \sum_{\bm{k}} \exp(- i \bm{k} \cdot \bm{L}) \; |i,\bm{k}\rangle .
\end{array}
\end{equation}
where $i$ is an orbital label, $\bm{L}$ is a crystal lattice vector, $N$ is the number of unit cells in the crystal, $\bm{k}$ is a wave vector in the crystal Brillouin-zone and $|i,\bm{k}\rangle = e^{-i{\bm k}\cdot{\bm r}} \ket{u_{i{\bm k}}}$ is a Bloch wave function
constructed from orbital $i$, and $|i,\bm{L}\rangle$ is a Wannier wavefunction.
We assume that the orbital identifications  
achieve maximally localized Wannier functions \cite{Wannier-1,Wannier-2} that can be physically identified with 
atomic orbitals on particular sites within the crystal's unit cell, or with particular chemical bonds.
Note that the Bloch functions $|i,\bm{k}\rangle$ are not energy eigenstates.
The construction of-maximally localized Wannier functions provides 
Hamiltonian representation that in most crystals is accurate and has a reasonably small matrix dimension and, 
importantly for what follows, provides a representation of the crystal's $\bm{k} \cdot \bm{p}$ Hamiltonian that is 
independent of momentum $\bm{k}$.  

Our transport theory assumes that the Wannier representation perfect-crystal Hamiltonian:
\begin{equation}\label{HWann}
H_0 = \sum_{{\bm L}{\bm L}' i i'} H_{{\bm L}{\bm L}'}^{i i'} \, \ket {i, {\bm L}} \bra{i', {\bm L}'}
\end{equation}
is known.  Translational symmetry guarantees that $H_{{\bm L}{\bm L}'}^{i i'}$ depends only on ${\bm L} - {\bm L}'$ for any $i$, $i'$.  The disorder models we discuss below assume that 
the real-space Wannier functions $ \langle \bm r \ket {i, {\bm L}}$ is localized near lattice 
site ${\bm L}$.  It follows from translational symmetry that the band Hamiltonian is diagonal in 
crystal momentum:
\begin{equation}
H_0 = \sum_{{\bm k}\, i'i} H_{\bm{k}}^{i i'}, \ket {i, {\bm k}} \bra{i', {\bm k}},
\end{equation} 
where 
\begin{equation} 
H_{\bm{k}}^{i i'} = \sum_{\bm L}  H_{{\bm L}{0}}^{i i'} \,  \exp( -i \bm{k} \cdot \bm{L}).
\end{equation} 
The band eigenstates $\varepsilon_{{\bm k}}^m$ are the eigenvalues of $H_{\bm{k}}^{i i'}$, so that 
\begin{equation}
H_0 = \sum_{{\bm k} m} \varepsilon_{{\bm k}}^m \, \ket {m, {\bm k}} \bra{m, {\bm k}},
\end{equation}
where $m$ is a band index, 
\begin{equation}\label{diag}
|m,\bm{k}\rangle = \sum_{i} z_i^{(m)} |i,\bm{k}\rangle
\end{equation}
and $z_i^{(m)}$ is the $m$-th eigenvector of $H_{\bm{k}}^{i i'}$. We refer to the representation provided by the $\{ |m,\bm{k}\rangle \}$ basis as the eigenstate representation, and write $|m,\bm{k}\rangle = e^{-i{\bm k}\cdot{\bm r}} \ket{u_{m{\bm k}}}$. Our formalism is designed to calculate the response to a steady electric field of the single-particle density matrix, which can be expressed in either the Wannier or the eigenstate representations: 
\begin{equation}
\arraycolsep 0.3ex
\begin{array}{rl}
\displaystyle \rho = & \displaystyle \sum_{{\bm L}{\bm L}' i i'} \rho_{{\bm L}{\bm L}'}^{ii'} \, \ket {i {\bm L}} \bra{i'{\bm L}'} \equiv \sum_{{\bm k}{\bm k}' m m'} \rho_{{\bm k}{\bm k}'}^{m m'} \, \ket {m {\bm k}} \bra{m' {\bm k}'}.
\end{array}
\end{equation}
These two representations will be used interchangeably throughout this work. 

\section{Disorder Models}
\label{Disorder}

With the exception of special cases, such as a clean undoped Dirac cone, response to a steady electric field is finite only in a disordered crystal. The transport formalism used in this paper is based on a Born approximation for the disorder potential. We comment below on the degree to which it can be generalized to stronger potentials by making a t-matrix expansion. As we see below, disorder then enters through averages of products of two eigenstate-representation disorder matrix elements of the following form; $\langle n_1',\bm{k}'|U|n_1,\bm{k}\rangle \;  \langle n_2',\bm{k}|U|n_2,\bm{k}'\rangle$. The average of the disorder potential is incorporated into the band Hamiltonian so that disorder averages of a single matrix element, $\overline{\langle n',\bm{k}'|U|n,\bm{k}\rangle }$, vanish by definition. We will discuss two disorder models: a random potential model and one based on random uncorrelated impurities. 

\subsection{Random potential}

Using the relationship between Bloch and Wannier basis functions we consider 
\begin{widetext}
\begin{equation}
\overline{ \langle i_{1'},\bm{k}'|U|i_1,\bm{k}\rangle \;  \langle i_{2'},\bm{k}|U|i_2,\bm{k}'\rangle } = \frac{1}{N^2}  \sum_{\bm{L}_i}  \exp[ i \bm{k} \cdot (\bm{L}_1-\bm{L}_{2'})] \;  \exp[ -i \bm{k}' \cdot (\bm{L}_{1'}-\bm{L}_{2})] \; \overline{ \langle i_{1'},\bm{L}_{1'}|U|i_1,\bm{L}_1\rangle \;  \langle i_{2'},\bm{L}_{2'} |U|i_2,\bm{L}_2 \rangle }
\end{equation}
We assume that the macroscopic response is identical for all disorder potentials,
allowing us to disorder average, and that translational symmetry is recovered after performing 
this average, implied above by the overline.  It follows that 
\begin{equation}
\overline{ \langle i_{1'},\bm{L}_{1'}|U|i_1,\bm{L}_1\rangle \;  \langle i_{2'},\bm{L}_{2'} |U|i_2,\bm{L}_2 \rangle }
= \overline{ \langle i_{1'},\bm{L}_{1'}+\bm{L} |U|i_1,\bm{L}_1 +\bm{L} \rangle \;  \langle i_{2'},\bm{L}_{2'} +\bm{L} |U|i_2,\bm{L}_2 +\bm{L} \rangle}
\end{equation}
for any $\bm{L}$.  
In other words we can set one of the four lattice vectors to zero in specifying independent disorder averages. 
Interpreting the Wannier-representation matrix elements using a tight-binding model language, 
matrix elements with $L_{i} \ne L_{i'}$ represent disorder in the hopping Hamiltonian while ones 
with $L_{i} = L_{i'}$ represent disorder in the atomic Hamiltonian on a particular site. 
The character of the disorder in a particular system is frequently not understood well enough to 
prefer one disorder model over another.  For the sake of definiteness, we assume
atomic disorder, and therefore set $\bm{L}_{1'} = \bm{L}_1$ and $\bm{L}_{2'} = \bm{L}_2$, obtaining
\begin{eqnarray}
\overline{ \langle i_{1'},\bm{k}'|U|i_1,\bm{k}\rangle \;  \langle i_{2'},\bm{k}|U|i_2,\bm{k}'\rangle }
= \frac{1}{N^2}  \sum_{\bm{L}_1,\bm{L}_2}  \exp[ i (\bm{k}-\bm{k}') \cdot (\bm{L}_1-\bm{L}_{2})] \overline{ \langle i_{1'},\bm{L}_{1}|U|i_1,\bm{L}_1\rangle \;  \langle i_{2'},\bm{L}_{2} |U|i_2,\bm{L}_2 \rangle }
\end{eqnarray}
Using translational symmetry this disorder average depends only on $\bm{L}_{1}-\bm{L}_{2}$.  
It follows that 
\begin{eqnarray}
\overline{ \langle i_{1'},\bm{k}'|U|i_1,\bm{k}\rangle \;  \langle i_{2'},\bm{k}|U|i_2,\bm{k}'\rangle } = \frac{1}{N} \sum_{\bm{L}}  \exp[ i (\bm{k}-\bm{k}') \cdot \bm{L}]  \;
\overline{ \langle i_{1'},\bm{L}|U|i_1,\bm{L}\rangle \;  \langle i_{2'},\bm{L}=0 |U|i_2,\bm{L}=0 \rangle }.
\end{eqnarray}
\end{widetext}
This is the most general disorder potential model we will consider.  The model can be simplified by assuming that 
there are no spatial correlations in the disorder potential.  In this case the disorder average reads
\begin{align}
&\overline{ \langle i_{1'},\bm{k}'|U|i_1,\bm{k}\rangle \;  \langle i_{2'},\bm{k}|U|i_2,\bm{k}'\rangle } \nonumber\\
&= \frac{1}{N}  \; \overline{ \langle i_{1'},\bm{L}|U|i_1,\bm{L}\rangle \;  \langle i_{2'},\bm{L} |U|i_2,\bm{L} \rangle }
\end{align}
is independent of wavevector. In the simplest model the disorder fluctuations in the on-site \textit{atomic-like} Hamiltonian are rigid energy shifts,
leading in the Wannier representation to 
\begin{equation}\label{U0W}
 \overline{ \langle i_{1'},\bm{k}'|U|i_1,\bm{k}\rangle \;  \langle i_{2'},\bm{k}|U|i_2,\bm{k}'\rangle } =  \displaystyle \frac{U_0^2}{N} \; \delta_{i_{1'},i_1} \delta_{i_{2'},i_2}, 
\end{equation} 
where $U_0$ typically has dimensions of energy $\times$ volume. In the Bloch representation the matrix elements reads
\begin{align}\label{U0B} 
&\overline{ \langle m_{1'},\bm{k}'|U|m_1,\bm{k}\rangle \;  \langle m_{2'},\bm{k}|U|m_2,\bm{k}'\rangle } \nonumber\\
&=  \displaystyle \frac{U_0^2}{N}  \;   \sum_{i_1} \; z^{(m_{1'})*}_{i_1} z^{(m_1)}_{i_1} \;  \sum_{i_2} \; z^{(m_{2'})*}_{i_2} z^{(m_2)}_{i_2}.
\end{align}
The simplified model of Eqs.~(\ref{U0W}) and (\ref{U0B}) can account for stronger atomic disorder potentials simply by viewing the disorder potential as a scattering matrix, but does 
not account for interference between scattering at different impurity sites. If the random potential is viewed as originating from a collection of impurities, interference between scattering processes at different impurity locations is present in the model provided one goes beyond the Born approximation to higher order in the scattering potential ($U^4$ and higher orders). In models that do not account for arbitrary band structures, the physics introduced above is often expressed in terms of a continuous disorder model in real space in which the expectation values of the various moments of the potential $U$ are specified as a function of the position ${\bm r}$. 
In the simplest version of Gaussian white noise, corresponding to Eqs.~(\ref{U0W}) and (\ref{U0B}), one typically writes $\bkt{U({\bm r}) U({\bm r}')} = (U_0^2/V) \delta({\bm r} - {\bm r}')$, where $V$ is the volume. Henceforth we shall use the abbreviated notation $U^{mm'}_{{\bm k}{\bm k}'} \equiv \langle m, \bm{k}| U |m', \bm{k}' \rangle$, for disorder potential matrix elements in a band eigenstate representation. The transport theory outlined in the following section, accounts for disorder only in terms of 
averages of the product of two disorder matrix elements.

\subsection{Uncorrelated impurities}

A related model regards the disorder potential as arising from a series of randomly distributed impurities: 
\begin{equation}
U ({\bm r}) = \sum_{\bm R} \mathcal{U} ({\bm r} - {\bm R}),
\end{equation}
where $\mathcal{U}$ represents the potential of a single impurity, and ${\bm R}$ labels the impurity locations. In the Bloch representation we have
\begin{equation}
U^{mm'}_{{\bm k}{\bm k}'} = \mathcal{U}^{mm'}_{{\bm k}{\bm k}'} \sum_{\bm R} e^{- i({\bm k} - {\bm k}')\cdot{\bm R}}.
\end{equation}
Since the impurity locations are random the sum vanishes. For the second-order term in the potential it is straightforward to prove that $\bkt{U^{mm'}_{{\bm k}{\bm k}'} U^{m''m'''}_{{\bm k}'{\bm k}}} = n_i \mathcal{U}^{mm'}_{{\bm k}{\bm k}'} \mathcal{U}^{m''m'''}_{{\bm k}'{\bm k}}$, where $n_i$ represents the impurity density. These expressions can be converted to the Wannier representation using the prescription outlined in Sec.~\ref{BW}, Eq. (\ref{diag}).

The expansion in powers of $n_i$ can be continued to any desired order. At order $U^4$, for example, interference between scattering processes at different impurity locations ${\bm R}$, ${\bm R}'$ will appear explicitly. 

\subsection{Disorder average}

The preceding discussion illustrates the fact that the expansion of the density matrix in the strength of the disorder potential can be formulated either in powers of $\tau$ (random potential model) or in powers of the impurity density $n_i$ (uncorrelated impurity model). In the remainder of this work we choose the latter for concreteness. In either case the disorder average can be constructed so as to capture the physics of the problem under study. For example, when repeated scatterings off the same impurity are known to be important one can replace the potential $U$ with the $t$-matrix. In this manner highly complex physics can be accounted for using our formalism, such as the Kondo effect \cite{Wang2013}. 

\section{Density-Matrix Kinetic Equation}
\label{DM}

The system is described by a single-particle density matrix $\rho$, which obeys the quantum Liouville equation 
\begin{equation}
\td{\rho}{t} + \frac{i}{\hbar} \, [H, \rho] = 0,
\end{equation}
where $H$ is the total Hamiltonian of the system. It is convenient to decompose the density matrix into two parts: one part, denoted by $\bkt{\rho}$, is averaged over impurity configurations, while the remainder, which is eventually integrated over, is denoted by $g$: 
\begin{equation}
\rho = \bkt{\rho} + g.
\end{equation}
In this section we will use angle brackets for disorder averages. Note that this definition implies that $\bkt{g}=0$. The Hamiltonian itself is decomposed into a band Hamiltonian, a disorder potential, and a perturbation $H_E$ due to the electric field, $H = H_0 + U + H_E$. 

\subsection{Scattering term in the Born Approximation}

The Born approximation is made to simplify the scattering term, the disorder contribution to the quantum Liouville equation. We now derive an expression for the scattering term which is valid in the absence of an external electric field by setting $H_E \to 0$. To this end we express the the quantum Liouville equation in terms of coupled equations for $\bkt{\rho}$ and $g$ as follows:
\begin{equation}
\arraycolsep 0.3ex
\begin{array}{rl}
&\displaystyle \td{\bkt{\rho}}{t} + \frac{i}{\hbar} \, [H_0, \bkt{\rho}] + \frac{i}{\hbar} \, \bkt{[U, g]} = 0, \\ [3ex]
&\displaystyle \td{g}{t} + \frac{i}{\hbar} \, [H_0, g] + \frac{i}{\hbar} \, [U, g] - \frac{i}{\hbar} \, \bkt{[U, g]} = - \frac{i}{\hbar} \, [U, \bkt{\rho}].
\end{array}
\label{coupled}
\end{equation}

In the Born approximation we can ignore the last two terms on the left hand side of the equation for $g$ because 
they contribute only beyond leading order in the disorder potential.  
What remains is a first-order in time inhomogeneous linear differential equation for $g$ which can be formally integrated
yield
\begin{equation}
g(t) = - \frac{i}{\hbar} \, \int^\infty_0 dt' \, e^{-iH_0t'/\hbar}[U, \bkt{\rho(t - t')}] e^{iH_0t'/\hbar}.
\end{equation}
In the Born approximation we can use the 
disorder-free expression for the time-evolution of the density matrix: 
$\bkt{\rho(t - t')} = e^{iH_0t'/\hbar}\bkt{\rho(t)}e^{-iH_0t'/\hbar}$ to obtain 
\begin{equation}
g(t) = - \frac{i}{\hbar} \, \int^\infty_0 dt' \, [e^{-iH_0t'/\hbar}U e^{iH_0t'/\hbar}, \bkt{\rho(t)}].
\label{goft}
\end{equation}
Note that we will no longer write the time dependence of $\bkt{\rho}(t)$ explicitly.
We obtain the full kinetic equation by substituting 
Eq.~(\ref{goft}) for $g$ in the first of Eqs.~(\ref{coupled}): 
\begin{equation}\label{kineq}
\arraycolsep 0.3ex
\begin{array}{rl}
\displaystyle \td{\bkt{\rho}}{t} + \frac{i}{\hbar} \, [H_0, \bkt{\rho}] + J(\bkt{\rho}) = 0,
\end{array}
\end{equation}
where the Born approximation scattering term is 
\begin{equation}
J(\bkt{\rho}) = \frac{1}{\hbar^2} \, \int^\infty_0 dt' \, \bkt{[U, [e^{-iH_0t'/\hbar}U e^{iH_0t'/\hbar}, \bkt{\rho(t)}]]}.
\end{equation}

\subsection{Scattering term in the eigenstate representation}

We now take advantage of the translational symmetry recovered after impurity-averaging 
by working in momentum space where $\bkt{\rho}$ is diagonal.  Because $J(\bkt{\rho})$ is diagonal and 
is the product of three matrices in the single-particle Hilbert space, one of which ($\bkt{\rho}$) is itself 
diagonal in momentum, the expression for $J(\bkt{\rho})$ at momentum $\bm{k}$ involves only one free 
intermediate momentum $\bm{k'}$: 
\begin{widetext}
\begin{equation}
\arraycolsep 0.3ex
\begin{array}{rl}
\displaystyle J_{\bm k}(\bkt{\rho}) = & \displaystyle \frac{1}{\hbar^2} \, \int^\infty_0 dt' \, \sum \{ \bkt{U^{mm'}_{{\bm k}{\bm k}'} U^{m'm''}_{{\bm k}'{\bm k}}} e^{-i(\varepsilon^{m'}_{{\bm k}'} - \varepsilon^{m''}_{{\bm k}}) t'/\hbar} \bkt{\rho}^{m''m'''}_{{\bm k}} - \bkt{U^{mm'}_{{\bm k}{\bm k}'} U^{m''m'''}_{{\bm k}'{\bm k}}} \bkt{\rho}^{m'm''}_{{\bm k}'} e^{-i(\varepsilon^{m''}_{{\bm k}'} - \varepsilon^{m'''}_{{\bm k}}) t'/\hbar} \\ [3ex]

& - \displaystyle \bkt{U^{mm'}_{{\bm k}{\bm k}'} U^{m''m'''}_{{\bm k}'{\bm k}}} e^{-i(\varepsilon^m_{\bm k} - \varepsilon^{m'}_{{\bm k}'})t'/\hbar} \bkt{\rho}^{m'm''}_{{\bm k}'}
+ \bkt{U^{m'm''}_{{\bm k}{\bm k}'} U^{m''m'''}_{{\bm k}'{\bm k}}} \bkt{\rho}^{mm'}_{\bm k} e^{-i(\varepsilon^{m'}_{\bm k} - \varepsilon^{m''}_{{\bm k}'})t'/\hbar} \}.
\end{array}\label{J-general-form}
\end{equation}
The final expression is obtained by regularizing the time integral by inserting a 
$e^{-\eta t'}$ convergence factor and using 
\begin{equation}
\frac{1}{\hbar^2} \int^\infty_0 dt' \, e^{-i(\varepsilon^{m'}_{{\bm k}'} - \varepsilon^{m''}_{{\bm k}} - i\eta) t'/\hbar} = \mathcal{P}\bigg[ \frac{1}{i\hbar(\varepsilon^{m'}_{{\bm k}'} - \varepsilon^{m''}_{{\bm k}})}  \bigg] + \frac{\pi}{\hbar} \, \delta(\varepsilon^{m'}_{{\bm k}'} - \varepsilon^{m''}_{{\bm k}}).
\end{equation}
The $\delta$-functions enforce energy conservation for real scattering events, 
while the principal part terms account for disorder-induced level repulsions. 
The complete expressions for these two contributions to the scattering term are listed in the Appendix \ref{Appendix-A}.

To solve the kinetic equation, we decompose the disorder-averaged part of the density matrix into two parts, $\bkt{\rho} = n + S$, where $n$ is diagonal in the band index, and $S$ is off-diagonal. Below we use the impurity density $n_i$ as a formal parameter to distinguish effects that arise at different orders in a disorder strength expansion. For the energy conserving scattering terms acting on the diagonal part of the density matrix we obtain
\begin{equation}\label{PJ}
\arraycolsep 0.3ex
\begin{array}{rl}
\displaystyle [J_d(n)]^{mm}_{{\bm k}} = & \displaystyle \frac{2\pi n_i}{\hbar} \sum_{m'{\bm k}'} \mathcal{U}^{mm'}_{{\bm k}{\bm k}'} \mathcal{U}^{m'm}_{{\bm k}'{\bm k}} \, (n^{mm}_{\bm k} - n^{m'm'}_{{\bm k}'}) \delta(\varepsilon^m_{\bm k} - \varepsilon^{m'}_{{\bm k}'}); \\ [3ex]
\displaystyle [J_{od} (n)]^{mm''}_{{\bm k}} = & \displaystyle \frac{\pi n_i}{\hbar} \sum_{m'{\bm k}'} \mathcal{U}^{mm'}_{{\bm k}{\bm k}'} \mathcal{U}^{m'm''}_{{\bm k}'{\bm k}} [ (n^{mm}_{\bm k} - n^{m'm'}_{{\bm k}'}) \delta(\varepsilon^m_{\bm k} - \varepsilon^{m'}_{{\bm k}'}) + (n^{m''m''}_{{\bm k}} - n^{m'm'}_{{\bm k}'}) \delta(\varepsilon^{m''}_{{\bm k}} - \varepsilon^{m'}_{{\bm k}'})],
\end{array}
\end{equation}
where the first equation is simply Fermi's Golden Rule and in the second equation it is understood that $m'' \ne m$. For the principal part terms
\begin{equation}\label{Principal}
\arraycolsep 0.3ex
\begin{array}{rl}
\displaystyle [J_{pp}(n)]^{mm''}_{{\bm k}} = & \displaystyle \frac{n_i}{i\hbar} \sum \mathcal{U}^{mm'}_{{\bm k}{\bm k}'} \mathcal{U}^{m'm''}_{{\bm k}'{\bm k}} \bigg[ (n^{m''m''}_{{\bm k}} - n^{m'm'}_{{\bm k}'}) \mathcal{P} \bigg(\frac{1}{\varepsilon^{m'}_{{\bm k}'} - \varepsilon^{m''}_{{\bm k}}}\bigg) + (n^{mm}_{\bm k} - n^{m'm'}_{{\bm k}'}) \mathcal{P} \bigg(\frac{1}{\varepsilon^{m}_{\bm k} - \varepsilon^{m'}_{{\bm k}'}} \bigg) \bigg].
\end{array}
\end{equation}
Note that the diagonal  ($m = m''$) elements of the principal term vanish and that the Born approximation collision term is linear in $\bkt{\rho}$.  We will not consider the principal-value
contributions to the collision terms further in this work, but note however that they can be incorporated into the general solution using the method outlined below for the elastic scattering terms. Finally, these expressions can be converted to the Wannier representation using Eq. (\ref{diag}).

\subsection{Beyond the Born approximation}

The expansion of Eqs.~(\ref{goft}) and (\ref{kineq}) can be continued to higher orders in the impurity potential. The leading-order contribution beyond the Born approximation the scattering term up to order $n_i$ is 
\begin{equation}
J(\bkt{\rho}) = - \frac{i}{\hbar^3} \, \int^\infty_0 dt' \, \bkt{[U, [e^{-iH_0t'/\hbar}U e^{iH_0t'/\hbar}, [e^{-iH_0t'/\hbar}U e^{iH_0t'/\hbar}, \bkt{\rho(t)}]]]}.
\end{equation}
Such an expansion is necessary when seeking to include for example skew scattering terms due to spin-dependent impurity potentials \cite{Bi2013}. 
For the case of side-jump terms the correct answer can be obtained in the Born approximation provided the electric field term $H_E$ is included in the time-evolution operator \cite{Culcer2010}. 
These specific matters will be discussed in a future publication.
 
We do expect that effects not included at the Born approximation level will sometimes be observable. One well known example is weak localization, which can be identified experimentally by characteristic field and temperature dependences. Additionally, it has been pointed out recently \cite{Ado2016} that interference between scattering processes at different impurity locations, not included in the Born approximation, can sometimes play an important role in Hall effects. We judge that the transport theory outlined above makes a good compromise between wide (but not universal) applicability and practicality for applications with realistic band structures.   

\end{widetext}

\section{Linear Response Theory}
\label{LRT}

\subsection{Numerical implementation procedure}

We are interested in the non-equilibrium expectation values of quantum mechanical observables, which are found by tracing the relevant operators with the density matrix. In the general case an operator may have a nonzero expectation value in equilibrium, although for common observables such as the charge current and spin polarization in paramagnetic systems the equilibrium expectation value is zero.

The Hamiltonian is assumed given in a representation of maximally localized Wannier functions as in Eq.~(\ref{HWann}). Expectation values may be evaluated either directly in the Bloch representation or using the Wannier representation. In the Bloch representation 
\begin{equation}\label{BlochExp}
{\rm Tr} (\hat{\rho}\hat{O}) = \sum_{\bm{k},n,n'} \, O^{nn'}_{\bm k} \, \rho^{n'n}_{\bm k}.  
\end{equation}
One resorts to Eqs.~(\ref{BlochWann}) and (\ref{diag}) to transform the matrix elements of $\hat{O}$ to the Bloch representation, while the matrix elements of the density matrix are given in Eqs.~(\ref{f0}), (\ref{rhoEdAllen}), (\ref{D0}) and (\ref{rhood}). 

In the Wannier representation
\begin{equation}
{\rm Tr} (\hat{\rho}\hat{O}) = \sum_{\bm{L}, \bm{L}',i,i'} \, O^{ii'}_{{\bm L}{\bm L}'} \, \rho^{i'i}_{{\bm L}'{\bm L}}.  
\end{equation}
The matrix elements $O^{ii'}_{{\bm L}{\bm L}'}$ in the Wannier representation are assumed known, whereupon one can use Eqs.~(\ref{BlochWann}) and (\ref{diag}) to convert the matrix elements given in Eqs.~(\ref{f0}), (\ref{rhoEdAllen}), (\ref{D0}) and (\ref{rhood}) to the Wannier representation. 

We note that Eq.~\ref{BlochExp} as written applies as long as the operator $\hat{O}$ is diagonal in wave vector. If one is seeking to calculate the charge or spin current expectation values in systems in which spin-orbit terms in the \textit{scattering} potential are important (e.g. skew scattering and side-jump contributions), the current operator in general contains terms off-diagonal in wave vector. In such cases Eq.~\ref{BlochExp} must be generalized to include a sum over ${\bm k}$-off-diagonal matrix elements of both $\hat{O}$ and $\hat{\rho}$.

\subsection{Driving term and kinetic equation}

When a constant uniform electric field is applied the term $H_E = e{\bm E}\cdot \hat{\bm r}$ is added to the Hamiltonian.
Below we assume that because ${\bm E}$ is small, ${\bm r}$ may be replaced 
by lattice vector ${\bm L}$ in the Wannier function Hamiltonian.  
The electric field therefore produces only a lattice-site dependent energy shift.
The quantum Liouville equation takes the form
\begin{equation}
\td{\rho}{t} + \frac{i}{\hbar} \, [H_0 + U, \rho] = - \frac{i}{\hbar} \, [H_E, \rho].
\end{equation}
In linear response theory the density matrix is decomposed into equilibrium and 
response components by definining $\bkt{\rho} = \bkt{\rho_0} + \bkt{\rho_E}$, where $\bkt{\rho_E}$ is
the correction to the equilibrium density matrix $\bkt{\rho_0}$ to first order in the electric field. We shall use this notation (subscript $0$ for equilibrium, subscript $E$ for the electric-field correction, no subscript for the sum of the two) for all components of the density matrix, thus $n_{\bm k} = n_{0{\bm k}} + n_{E{\bm k}}$ and so forth. By linearizing the kinetic equation with respect to $E$ and noting that $J(\rho_0) = 0$, we find that 
\begin{equation}\label{kineq2}
\arraycolsep 0.3ex
\begin{array}{rl}
\displaystyle \td{\bkt{\rho_E}}{t} + \frac{i}{\hbar} \, [H_0, \bkt{\rho_E}] + J(\bkt{\rho_E}) = - \frac{i}{\hbar} \, [H_E, \bkt{\rho_0}].
\end{array}
\end{equation}
Because all three terms are on the left had side are linear in $\rho_{E}$, the linear response can be 
evaluated by performing a formal matrix inversion.  

We now discuss the form of the driving term on the right hand side of Eq.~(\ref{kineq2})
using the Wannier and Bloch eigenstate representations. In the Wannier representation the driving term takes the form
\begin{equation}
\arraycolsep 0.3ex
\begin{array}{rl}
\displaystyle - \frac{i}{\hbar} \, \tbkt{i, {\bm L}}{[H_E, \bkt{\rho_0}]}{i', {\bm L}'} = & \displaystyle - \frac{ie{\bm E} \cdot ({\bm L} - {\bm L}')}{\hbar} \, \bkt{\rho_0}_{{\bm L}{\bm L}'}^{ii'}.
\end{array}
\end{equation}
In the Bloch representation the equilibrium density matrix is given by  
\begin{equation}\label{f0} 
\tbkt{m, {\bm k}}{\bkt{\rho_0}}{m', {\bm k}} = f_0(\varepsilon^m_{\bm k}) \delta^{mm'} \equiv n_{0{\bm k}}^{mm},
\end{equation}
where $f_0(\varepsilon^m_{\bm k})$ is the Fermi-Dirac distribution function evaluated at energy $\varepsilon^m_{\bm k}$. To obtain the driving term we substitute $H_E = e{\bm E}\cdot \hat{\bm r}$ into the commutator $[H_E, \bkt{\rho_0}]$, and use the fact that $|m,\bm{k}\rangle = e^{-i{\bm k}\cdot{\bm r}} \ket{u_{m{\bm k}}}$ to re-express terms of the form
\begin{equation}
\hat{\bm r} |m,\bm{k}\rangle = \bigg[ i \, \pd{}{\bm k} \, e^{-i{\bm k}\cdot{\bm r}}\bigg] \ket{u_{m{\bm k}}}.
\end{equation}
With this substitution the driving term can immediately be written in the form \cite{Comment-A}
\begin{align}\label{DEk}
&- \frac{i}{\hbar} \, \tbkt{m, {\bm k}}{[H_E, \bkt{\rho_0}]}{m', {\bm k}} = \nonumber\\
 & \frac{e{\bm E}}{\hbar} \cdot \bigg\{\delta^{mm'} \frac{\partial f_0(\varepsilon^m_{\bm k})}{\partial {\bm k}}+i \bm{\mathcal R}_{\bm k}^{mm'} [f_0(\varepsilon^m_{\bm k}) - f_0(\varepsilon^{m'}_{\bm k})]\bigg\},
\end{align}
where
\begin{equation}
\bm{\mathcal R}_{\bm k}^{mm'} = \dbkt{u^m_{\bm k}}{i \pd{u^{m'}_{\bm k}}{\bm k}}
\end{equation}
is a momentum space Berry (gauge) connection. This is the part of the driving term that gives rise to the momentum-space Berry curvature intrinsic contribution to the Hall conductivity of systems with broken time reversal symmetry, and also to other response properties in other systems. The Fermi occupation number difference factor $f_0(\varepsilon^m_{\bm k}) - f_0(\varepsilon^{m'}_{\bm k})$ makes it evident that the term in square brackets drives off-diagonal response, $m \ne m'$, and therefore inter-band coherence contributions to the electrical response of the solid. Equation (\ref{DEk}) specifies the full intrinsic driving term, which is determined only by the system's electronic structure.

\subsection{Diagonal part of the density matrix to leading order in the impurity density}
\label{sec:diag}

Using Eq.\ (\ref{DEk}) and the Born approximation for the collision term, the density matrix response can be organized as in an expansion in powers of $n_i$.  Because $[H_0, \bkt{\rho_E}]$ is purely off-diagonal in the band index, the leading steady-state (time-independent) response is purely diagonal and proportional to $n_i^{-1}$, hence it is denoted by $n_{E{\bm k}}^{(-1)}$:
\begin{equation}\label{KineticEq_sim}
J_d[n_{E}^{(-1)}]^{mm}_{\bm k} = \frac{e{\bm E}}{\hbar} \cdot \pd{f_0(\varepsilon^m_{\bm k})}{{\bm k}},
\end{equation}
The collision contribution to the kinetic equation can be viewed as a matrix operator acting on the density matrix.
Considering for the moment only the diagonal response we define the collision matrix $K$ by  
\begin{equation} 
J_d(n_E)^{mm}_{{\bm k}} = n_i \, \sum_{m'{\bm k'}} \, K_{m{\bm k},m'{\bm k'}} \, n^{m'm'}_{E{\bm k'}}.
\end{equation}
$K^d$ acts on the density-matrix considered as a vector with components labelled by band and wavevector.
Comparing with Eq.~(\ref{PJ}) we have that 
\begin{align}
K_{m{\bm k},m'{\bm k'}} =& \frac{2\pi n_i}{\hbar} \Big[ \delta_{m,m'} \delta_{{\bm k},{\bm k'}}  \sum_{m''{\bm k}''} \mathcal{U}^{mm''}_{{\bm k}{\bm k}''} \mathcal{U}^{m''m}_{{\bm k}''{\bm k}} - \mathcal{U}^{mm'}_{{\bm k}{\bm k}'} \mathcal{U}^{m'm}_{{\bm k}'{\bm k}} \Big] \nonumber\\
&\times\delta(\varepsilon^m_{\bm k} - \varepsilon^{m'}_{{\bm k}'}).
\end{align}
Solving for $n_{E{\bm k}}$ to leading order in $n_i$ we find that \cite{Allen_PRB78}
\begin{equation}\label{rhoEdAllen}
n_{E{\bm k}}^{(-1), mm} = n_i^{-1} \, e{\bm E} \sum_{m'{\bm k}'} K^{-1}_{m{\bm k},m'{\bm k'}}  \cdot {\bm v}^{m'}_{\bm k'} \, \pd{f_0(\varepsilon^{m'}_{\bm k'})}{\varepsilon^{m'}_{\bm k'}}.
\end{equation}
where we have used $ {\bm v}^m_{\bm k} = (1/\hbar) \, (\partial{\varepsilon^m_{\bm k}}/\partial{\bm k})$.  Note that 
the total density response summed over all bands and wavevectors is a zero eigenvalue of 
$K_{m{\bm k},m'{\bm k'}}$ so that  $K^{-1}_{m{\bm k},m'{\bm k'}}$ is strictly speaking undefined. 
However, the driving term does not act on this eigenvector.  We implicitly assume above that the 
matrices $K$ and $K^{-1}$ have been decomposed to remove the total density from the vector space.
In the simplest case, that of a metal, one may replace
\begin{equation}
\pd{f_0(\varepsilon^m_{\bm k})}{\varepsilon^m_{\bm k}} = - \delta(\varepsilon^m_{\bm k} - \varepsilon_F).
\end{equation}

\subsection{Off-diagonal part of the density matrix and anomalous driving term}

Because $[H_0, \bkt{\rho_E}]$ is non-zero off the diagonal, the linear response of off-diagonal
density- matrix components to a steady electric field, $S_E$, 
does not diverge in the absence of disorder.  The leading response is therefore independent of 
disorder strength and starts at the next-to-leading order in $n_i$, which we can explicitly label by $^{(0)}$:
\begin{equation}\label{rhoperp}
\td{S_{E{\bm k}}^{(0), mm'}}{t} + \frac{i}{\hbar} \, [H_{0{\bm k}}, S_{E{\bm k}}^{(0)}]^{mm'} = D^{mm'}_{E{\bm k}} + D'^{\, mm'}_{E{\bm k}}.
\end{equation}
The right hand side in Eq.~(\ref{rhoperp}) has two contributions. The first is the intrinsic off-diagonal driving term, 
\begin{equation}\label{intdr}
D^{mm'}_{E{\bm k}} = - \frac{e{\bm E}}{\hbar} \cdot \dbkt{u^m_{\bm k}}{\pd{u^{m'}_{\bm k}}{\bm k}} [f_0(\varepsilon^m_{\bm k}) - f_0(\varepsilon^{m'}_{\bm k})],
\end{equation}
which represents the Fermi sea response and contains all the Berry phase like contributions. 
The second depends on disorder and we refer to it below as the anomalous driving term 
$D'^{\, mm'}_{E{\bm k}} = - J_{od}[n_{E{\bm k}}^{(-1)}]^{mm'}$ which 
is due to scattering and is non-zero only on the Fermi surface. 
This extrinsic term will be written in a more explicit form in Eq. (\ref{Dprime}). 
The solution to Eq. (\ref{rhoperp}) is straightforwardly found
\begin{equation}
S_{E{\bm k}}^{(0)} =  \int_0^\infty dt' \, e^{-iH_{0{\bm k}}t'/\hbar}  (D_{E{\bm k}} + D'_{E{\bm k}}) e^{iH_{0{\bm k}}t'/\hbar}.
\end{equation}
It is regularized by inserting an infinitesimal $e^{-\eta t'}$ and taking the limit $\eta \rightarrow 0$ after integrating over time
\begin{align}\label{rhood}
S_{E{\bm k}}^{(0), mm'} =& - i\hbar \mathcal{P} \bigg( \frac{D^{mm'}_{E{\bm k}} + D'^{\, mm'}_{E{\bm k}} }{\varepsilon^m_{\bm k} - \varepsilon^{m'}_{\bm k}} \bigg)\nonumber\\
& + \pi \hbar (D^{mm'}_{E{\bm k}} + D'^{\, mm'}_{E{\bm k}} )  \,\delta(\varepsilon^m_{\bm k} - \varepsilon^{m'}_{\bm k}).
\end{align}
In Eq. (\ref{rhood}) we have separated the principal part and the $\delta$-function terms in the time integral.
The $\delta$-function terms are important near points in momentum space where 
different bands touch , as discussed below. 
The matrix-elements of $J_{od}[n_{E{\bm k}}^{(-1)}]^{mm'}$ can be identified 
from Eq.\ (\ref{PJ}). 
Using Eq. (\ref{PJ}), we express explicitly the anomalous driving term
$D'^{\, mm''}_{E{\bm k}} = - J_{od}[n_{E{\bm k}}^{(-1)}]^{mm''}$ as
\begin{align}\label{Dprime}
D'^{\, mm''}_{E{\bm k}} = & \displaystyle  \frac{\pi n_i}{\hbar} 
 \sum_{m'{\bm k}'} \mathcal{U}^{mm'}_{{\bm k}{\bm k}'} \mathcal{U}^{m'm''}_{{\bm k}'{\bm k}} \nonumber\\
 &\times \left\{ [
n_{E{\bm k}}^{(-1), mm} 
 -  n_{E{\bm k'}}^{(-1), m'm'} ] \delta(\varepsilon^m_{\bm k} - \varepsilon^{m'}_{{\bm k}'})\right. \nonumber\\
&\left. + \displaystyle [ n_{E{\bm k}}^{(-1), m''m''}  -  n_{E{\bm k}}^{(-1), m'm'}] 
\delta(\varepsilon^{m''}_{{\bm k}} - \varepsilon^{m'}_{{\bm k}'}) \right\}.
\end{align}
Using Eqs.\ (\ref{intdr}), (\ref{rhood}) and (\ref{Dprime}) the full off-diagonal response of the crystal can be determined to leading order in the disorder potential. 

\subsection{Diagonal part of the density matrix to next-to-leading order in the impurity density}
\label{sec:subdominant}

The sub-leading order contribution to the diagonal density matrix, 
distinguished by the superscript $^{(0)}$ satisfies the equation
\begin{equation}\label{D0}
J_d[n_{E}^{(0)}]^{mm}_{\bm k} = - J_d[S_{E}^{(0)}]^{mm}_{\bm k}.
\end{equation}
where $S_{E}^{(0)} \equiv S_{E{\bm k}}^{(0)}$ is found using Eq.\ (\ref{rhood}). The RHS of Eq.~(\ref{D0}) plays the role of a supplementary effective diagonal driving term that arises from the collision kernel acting on the off-diagonal response. The RHS plays the role of a driving term and the LHS needs to be solved. This equation can be solved for $n_{E{\bm k}}^{(0)}$ using the method explained above, i.e., by letting $K^{-1}$ act on the effective driving term, as inf Ref.~\onlinecite{Allen_PRB78}. The expansion now contains all terms in the density matrix up to order zero in the impurity density (disorder strength), since the next iteration of $S_{E{\bm k}}$ will be order $^{(1)}$ i.e. $\propto n_i$. 

\section{Discussion}
\label{Disc}

Our transport theory addresses the influence of a constant electric field on the density-matrix of 
independent electrons in a weakly disordered crystal.  In real crystals, electron-electron 
interactions always play an essential role, even when electrons form a Fermi liquid state.
In practice we imagine the independent fermions that appear in our theory as
the quasiparticles of a mean-field-like description of a particular crystal of interest.
The Kohn-Sham quasiparticles of density-functional-theory (DFT) are of particular interest 
because DFT combined with an appropriate exchange-correlation energy functional
provides a {\it good enough} description of many important solid state systems. Because the
Kohn-Sham quasiparticles are defined by an effective Hamiltonian that depends on electron
charge and spin- or current- density functions, which can themselves be altered in an important way 
by an applied electric field, applications of the transport theory might in many instances require 
that the response be self-consistently incorporated in the Kohn Sham Hamiltonian.  
This is particularly true in the case of solid state systems with order that can 
be manipulated magnetically, for example the many interesting effects 
that are important in spintronics.   

\subsection{Band-diagonal and off-diagonal contributions}

The formulation of transport theory summarized above demonstrates that, to sub-leading order in disorder strength, the linear response of observables to a steady electric field depends in general on the interplay between three contributions distinguished by the way in which they contribute to the single-particle density matrix in the
band-eigenstate representation: (i) a contribution to the band-diagonal part of the density matrix which balances the intra-band driving term and scattering on the Fermi surface, and diverges in the limit of very weak disorder scattering; (ii) a contribution to the band off-diagonal part of the density matrix that is independent of disorder character and 
hence an intrinsic band-structure property involving the entire Fermi sea and (iii) an extrinsic contribution to both the band off-diagonal and the band-diagonal parts of the density matrix that is finite in the weak disorder limit and originates from the collision kernel acting on the leading Fermi surface response (i). Given a Wannier interpolation of a crystal Hamiltonian it is practical to fully and accurately solve the transport equations we have derived using modern computational resources.  

The leading band-diagonal response [(i) above] of electrons close to the Fermi surface is the subject of most transport theories in metals \cite{Allen_PRB78} and, indeed, it normally dominates the response of most observables, longitudinal current in particular, in reasonably good metals. It has been studied extensively in the past and is well understood \cite{Allen_PRB78}.

\subsection{Contributions to inter-band coherence}
 
The importance for some observables of the intrinsic off-diagonal driving term $D_{E{\bm k}}$ in practical solid state materials 
under typical experimental conditions first came to light in recent theoretical efforts to achieve a 
quantitative understanding of the anomalous Hall effect, the Hall effect in the absence of a magnetic field,
which is non-zero in ferromagnetic and some non-collinear antiferromagnetic systems \cite{Jungwirth_PRL02, Nagaosa-AHE-2010, Sinitsyn_AHE_Review_JPCM08, Crepieux_AHE_KuboDirac_PRB01, AHE_vertex_PRL_2006, Sinitsyn_AHE_KuboStreda_PRB07, Kovalev_Multiband_AHE_PRB09, Nagaosa_AHE_PRB08, Culcer_AHE_PRB03, Chang_TIF_ferromagnetism_AHE_AM2013, Culcer_TI_AHE_PRB11, Nomura_PRL_2011, ZangNagaosa_TI_Monopole_PRB10}.
The intrinsic Hall current response is expressed exactly in terms of the momentum-space 
Berry curvature of the crystal bands, and can be non-zero even in insulators \cite{Jungwirth_PRL02, Culcer_AHE_PRB03, Sinitsyn_AHE_Review_JPCM08, Nagaosa-AHE-2010}.
In two-dimensional insulators, the Hall conductivity is proportional to the integral of the 
Berry curvature over the two-dimensional Brillouin-zone, which is a quantized topological index of the 
band structure, the first theoretically understood classification of crystal band structures \cite{TKNN}.
The quantum Hall effect of two-dimensional insulators is an example of an important 
electric field response of an insulator, which must originate entirely from the off-diagonal 
terms because of the absence of a Fermi surface. Insulator response to an electric field 
is also important in spintronics, which has witnessed an increasingly active role for magnetic insulators.  
In metals, the intrinsic response (ii) and the much more complex extrinsic sub-leading response (iii) must be 
treated on an equal footing in order to achieve reliable theoretical results.  
Our transport formalism is motivated by the desirability of meeting this need, not only in 
toy model systems, but also in crystals with complex electronic structure.  
The anomalous driving term $D'_{E{\bm k}}$ (iii) leads to a scattering correction to the intrinsic response. 
It arises from a projection of the collision kernel acting on the diagonal part of the density 
matrix $n_{E{\bm k}}$, which contains a $\delta$-function at the Fermi energy. 
Hence this contribution involves the carriers on the Fermi surface. 
This correction can be sizable in magnitude and in many 
cases cancels out the total response of observables of interest \cite{Inoue_RashbaSHE_Vertex_PRB04, Raimondi_SHE_PRB06, CulcerWinkler_NonSHE_PRL07, Culcer_SteadyState_PRB07}.

The expansion in disorder strength of the off-diagonal part of the density matrix starts at order $n_i^{(0)}$. To see this, note firstly that the intrinsic off-diagonal driving term $D_{E{\bm k}}$ has no dependence on disorder. Secondly, referring to Eq.\ (\ref{Dprime}), in the anomalous driving term $D'_{E{\bm k}}$ the factor of $n_i$ cancels the factor of $1/n_i$ contained in the scattering time $\tau_m$, making this term apparently independent of the impurity density. This is simply a reflection of the fact that (i) the non-equilibrium correction to the density matrix is an expansion in powers of $n_i$ and (ii) the leading term in the expansion is $\propto n_i^{-1}$, since it is linear in the transport scattering time that is 
needed to keep the Fermi surface near equilibrium. The next-to-leading term is thus of order $n_i^{(0)}$. (Alternatively, if one uses $U^2$ instead of $n_i$ to characterise the strength of the disorder potential, then this contribution to the density matrix appears to be independent of the magnitude of the disorder potential.)

It is interesting to consider briefly the limiting case in which inter-band disorder matrix elements happen to be much smaller than intra-band ones. In this case the inter-band coherence contribution is dominated by the intrinsic off-diagonal driving term, and the anomalous driving term is negligible. If, in addition, the diagonal disorder matrix elements happen to differ strongly between two bands then effectively the dynamics is dominated by one band. However, realistic disorder potentials, such as bare and screened Coulomb potentials and short-range disorder, as well as spin-dependent potentials leading to spin flip and skew scattering, always have off-diagonal matrix elements that are comparable in magnitude to the diagonal ones.

\subsection{Inter-band coherence effects in conductors and insulators}

In most conducting crystals whether metals, semiconductors or semimetals all three terms (i)-(iii) are present in the linear response to an electric field. In non-magnetic conductors with strong spin-orbit interactions the band-diagonal part of the density matrix is responsible for the steady-state spin density (alternatively the current-induced spin polarization) \cite{Edelstein_SSC90, Inoue_RashbaSHE_Vertex_PRB04, Culcer_CISP_PRB05, CulcerWinkler_NonSHE_PRL07, Culcer_SteadyState_PRB07, Culcer_TI_Kineq_PRB10}.
The band off-diagonal part is responsible for the spin-Hall effect, which typically has sizable contributions from both the intrinsic driving term (ii) and the anomalous driving term (iii) \cite{Raimondi_SHE_PRB06, CulcerWinkler_NonSHE_PRL07, Culcer_SteadyState_PRB07}. In massless Dirac fermion systems such as graphene and topological insulators the inter-band coherence terms (ii) and (iii) give rise to the well-known Zitterbewegung contribution to the minimum conductivity \cite{Katsnelson_Zbw_MinCond_EPJB06, Adam_Gfn_PNAS07, Mishchenko_EPL08, Culcer_Gfn_Transp_PRB08, CastroNeto2009, Culcer_Bil_PRB09, Trushin_BLG_MinCond_PRB10, SDS_Gfn_RMP11}. This contribution is traced to the $\delta$-function terms in Eq.~(\ref{rhood}), which are important at the Dirac point, when the bands touch. Hence the off-diagonal part of the density matrix contains a reactive response to electric fields and is associated with non-adiabatic corrections to the carrier motion. 

In magnetically-doped conductors the band off-diagonal part of the density matrix is responsible for the anomalous Hall effect, which in turn has strong contributions from both the intrinsic and anomalous driving terms \cite{AHE_vertex_PRL_2006, Sinitsyn_AHE_KuboStreda_PRB07, Nagaosa_AHE_PRB08, Sinitsyn_AHE_Review_JPCM08, Kovalev_Multiband_AHE_PRB09, Nagaosa-AHE-2010}. The extreme case of this is the quantum anomalous Hall effect occurring in topological insulators in which the chemical potential lies in the surface energy gap opened by the magnetization \cite{Chang_TIF_ferromagnetism_AHE_AM2013}. In that case the surface valence band is full while the surface conduction band is empty, hence the band-diagonal part of the density matrix is zero and the anomalous driving term likewise vanishes \cite{Culcer_TI_AHE_PRB11}. The quantized anomalous Hall response results from the intrinsic off-diagonal driving term $D_{E{\bm k}}$. A similarly complex interplay between the diagonal and off-diagonal response enters the calculation of spin-orbit torques in ferromagnetic structures (see below).

Finally, in insulators the diagonal driving term vanishes identically and consequently the anomalous off-diagonal driving term is also zero. The only possible response is due to the intrinsic off-diagonal driving term, which leads to e.g. the inter-band polarization. 

\subsection{Computational applications}

The aim of this work is to provide a general quantum kinetic equation that can be easily adapted to computational strategies to study inter-band coherence effects in systems with complex band structures that cannot be described by simple analytical models. In this subsection we discuss briefly a series of problems that are amenable to a treatment based on the quantum kinetic equation as presented above. 

A considerable amount of research has focused on the possibility of switching the magnetization of a ferromagnet by passing an electrical current through it. Initial studies focussed on spin-transfer torques, which exploited the inhomogeneity in the magnetization in the vicinity of an interface. Recent research has highlighted the role of spin-orbit torques in materials with strong spin-orbit interactions. The calculation of the spin-orbit torque is equivalent to finding the steady-state spin density induced by an external electric field, which lends itself to the treatment introduced in this work. Computational methods in this case are indispensable since the materials under study are typically metals with complicated band structures involving several bands with complex topologies intersecting the Fermi surface. 

An interesting recent development in this direction concerns interfacing ferromagnets with topological materials with strong spin-orbit coupling. Experimental work has shown that the steady-state spin density in a topological insulator gives rise to a sizable spin-orbit torque in an adjacent ferromagnet \cite{Mellnik2014}. 
In principle, for a topological insulator one can use a simple two-band Dirac model to determine the spin density given a certain disorder realization. Yet one recognized limitation of using topological insulators is that the spin density lies only in the plane containing the interface. This has led to the exploration of novel topological materials such as transition metal dichalcogenides. Among these the type-II Weyl semimetal WTe$_2$ has been shown to give rise to a strong out-of-plane spin-orbit torque due to the absence of mirror symmetry \cite{Xu2016,Das2016,Wang2016,MacNeill2017}. 
An accurate calculation of this torque can only be done numerically due to the complexities of the band structure of WTe$_2$. A simple analytical model of WTe$_2$ consists of a \textit{tilted} Dirac cone such that one band becomes effectively flat and the Fermi surface appears to be open. This unphysical feature is not present in the numerically calculated band structure, yet no accurate analytical model currently exists. In addition, the Fermi surface has been shown to be complex, with electron and hole pockets concomitantly present. 

Likewise the spin-Hall and inverse spin-Hall effect in metals continue to be the subject of a significant amount of work and controversy \cite{Sinova2015}. 
The kinetic equation has been shown to capture this effect accurately. Moreover, a numerical formulation can accommodate different boundary conditions with relative ease, a fact that is of the utmost importance in this problem, since the effect of the spin-Hall current can only be detected through the spin accumulation it generates at the boundary of the sample. The theoretical framework can be extended to cover spin-dependent scattering mechanisms leading to skew scattering and side jump, as well as to the correct definition of the conserved spin current \cite{Shi2006}, 
which is cumbersome to deal with analytically. 

As a final example, we note that the kinetic equation can be straightforwardly extended to describe magnetotransport in materials with arbitrary band structures, revealing the physical origin of complex phenomena involving both Berry phase and scattering effects, such as the chiral anomaly of Weyl semimetals. This will be done in a forthcoming publication \cite{Sekine2017}. 

\subsection{Comparison with equivalent formulations of linear response theory}

The formulation of linear response theory that we present converts the quantum Liouville equation into an effective semiclassical kinetic equation that is exactly equivalent to the quantum Boltzmann equation. It is very similar in spirit to the Keldysh method \cite{Comment-C}. In the same way as the quantum Boltzmann equation our approach can be generalised to inhomogeneous systems by taking the Wigner transform of the density matrix and making a gradient expansion in the spatial variable. However, the approach presented in this work is considerably more intuitive and transparent than the Keldysh method since the kinetic equation is formulated in terms of the density matrix, which can be directly associated with quantum mechanical expectation values. It avoids the cumbersome steps needed to convert the Keldysh Green's function into an effective distribution function, such as the necessity of an \textit{Ansatz} for the Keldysh component and integration over an additional energy variable, which become increasingly opaque in complex, multiband systems. 

The method we propose has a strong parallel with the Kubo approach based on the fluctuation-dissipation theorem. Whereas the starting point of both approaches is the quantum Liouville equation, the Kubo approach focuses on solving the quantum Liouville equation immediately in integral form and then expanding this solution in the strength of the disorder potential. The Kubo method may thus be termed the \textit{integral} approach to the kinetic equation, while the method we discuss could be viewed as the \textit{differential} approach. The propagators appearing as part of the Kubo method are equivalent to the time-evolution operators in our density-matrix language. The main methodological difference is in the way the disorder is treated: the Kubo approach essentially incorporates the disorder potential $U$ directly into the time-evolution operator, and then expands this in the strength of the disorder potential. The density-matrix approach builds up the disorder expansion order by order before solving for the density matrix using the time evolution operator for the clean system. The vertex correction due to the disorder ladder diagrams, which in the Kubo formalism leads to the correct expression for the momentum relaxation time, is equivalent to the scattering-in term in the kinetic equation, and is much easier to obtain in the language discussed in this work. It does not require summation over an infinite series of terms, which would be rather inconvenient for numerical implementations. 

All the results we have obtained using the density-matrix approach to date have matched those obtained using the Kubo formalism. This includes the steady-state spin density and spin current \cite{CulcerWinkler_NonSHE_PRL07, Culcer_SteadyState_PRB07,Winkler2008}, 
as shown below, 
the skew scattering and side-jump contributions to the spin-Hall effect both in the presence and in the absence of band structure spin-orbit coupling \cite{Culcer2010,Bi2013,TseSDS_SJ_PRL06,Raimondi2012}, 
the anomalous Hall effect \cite{Culcer_TI_AHE_PRB11}, 
the longitudinal conductivity of graphene and the minimum conductivities of intrinsic monolayer and bilayer graphene \cite{Culcer_Gfn_Transp_PRB08, Culcer_Bil_PRB09}. 

Aside from its intuitive transparency and straightforward physical interpretation, the advantage of the density-matrix method is that one can immediately separate intrinsic effects, extrinsic effects, and effects that combine interband coherence and disorder. For example, in the density-matrix language it is immediately obvious that the well-known vertex correction to the spin-Hall current represents the presence of a steady-state spin density \cite{CulcerWinkler_NonSHE_PRL07,Culcer_SteadyState_PRB07}. 
Building on this insight, the vanishing of this vertex correction to the spin-Hall effect in models such as the cubic Dresselhaus and spherical Luttinger models becomes self-explanatory, since a steady-state spin density in these non-gyrotropic systems is forbidden by symmetry. 

We stress that the choice of disorder model is independent of the approach used. The models we discussed above for the case of the quantum Liouville equation - a random disorder potential and uncorrelated impurities - are frequently used in conjunction with the Kubo formula and the Keldysh formulation. 

\section{Applications of the theory}
\label{Appl}

In this section we discuss the application of the general theory developed above to a few specific simple model examples, fully recovering results derived previously using a variety of 
different techniques.  Our intention in this section is to establish using a few specific examples, that our Born approximation theory contains all recognized physical effects.  It is however formulated in this paper in a way which is appropriate for arbitrarily complex band structures. 

For the sake of simplicity the focus of this section will be on spin-dependent effects in spin-orbit coupled semiconductors described by the Rashba model, including the electric-field induced spin polarization, the spin-Hall effect and the anomalous Hall effect. We stress that, very generally, whenever intrinsic inter-band coherence effects are strong, extrinsic effects are also expected to be important. This fact emerges explicitly in the examples given below. A related example, not studied in detail here but considered in a recent publication \cite{Culcer_TI_AHE_PRB11}, is that of the anomalous Hall effect in topological insulators doped with magnetic impurities. In this case, as shown in Ref.~\onlinecite{Culcer_TI_AHE_PRB11}, the intrinsic contributions due to the conduction and valence bands cancel each other out, leaving the extrinsic contribution due to the anomalous driving term as the \textit{only} sizable contribution to the anomalous Hall effect. This highlights the importance of studying intrinsic and extrinsic terms on the same footing.

\subsection{Linear Rashba model for a non-magnetic semiconductor}

The Hamiltonian for an inversion asymmetric two-dimensional electron gas is
\begin{equation}
H_{0{\bm k}} = \frac{\hbar^2 k^2}{2m^*} + \alpha (\sigma_x k_y - \sigma_y k_x) = \frac{\hbar^2 k^2}{2m^*} + \alpha k \begin{bmatrix} 0 & i e^{-i\theta} \cr -i e^{i\theta} & 0 \end{bmatrix},
\end{equation}
where $m^*$ is the electron effective mass, $\alpha$ the spin-orbit constant, $\sigma_i$ are the Pauli matrices, and $\theta$ the polar angle of the wave vector. The eigenvectors are
\begin{equation}\label{Rashba_eignve}
\ket{u_{\bm k}^\pm} = \frac{1}{\sqrt{2}} \, \begin{bmatrix} e^{-i\theta} \cr \mp i \end{bmatrix},
\end{equation}
and the energies are $\varepsilon_{\bm k}^\pm = \hbar^2 k^2/2m^* \pm \alpha k$. The two band indices here are $n = \pm$. We assume impurities to be short-ranged so that there is only one relaxation time for each band (Bloch lifetime = momentum relaxation time) which we call $\tau_\pm$. 

\subsubsection{Diagonal part of the density matrix and spin density}

The diagonal part of the density matrix at $T=0$ is
\begin{equation}\label{Rashbarhopm}
\arraycolsep 0.3ex
\begin{array}{rl}
\displaystyle n^{(-1),\pm\pm}_{E{\bm k}} = & \displaystyle - \frac{e{\bm E} \cdot\hat{\bm k} \tau_\pm}{\hbar} \, \bigg( \frac{\hbar^2k}{m^*} \pm \alpha \bigg) \, \delta(\varepsilon^\pm_{\bm k} - \varepsilon_F),
\end{array}
\end{equation}
where $\hat{\bm k}=(k_x/k, k_y/k)$ is a unit vector along the $\bm{k}$ direction. We consider a short-range (on-site) disorder of the form $U(\bm{r})=U_0\sum_i\delta(\bm{r}-\bm{r}_i)$, and assume that the correlation function satisfies $\langle U(\bm{r})U(\bm{r}')\rangle=n_{i}U_0^2\,\delta(\bm{r}-\bm{r}')$ with $n_{i}$ the impurity density. The relaxation times $\tau_\pm$ for short-range impurities are found through
\begin{align}
\frac{1}{\tau_\pm(k)} = \frac{n_i U_0^2}{\hbar} \, \int dk'\, k' \, \delta(\varepsilon^\pm_{{\bm k}'} - \varepsilon^\pm_{\bm k})
 = \frac{n_i m^* U_0^2}{\hbar^3} \, \bigg(\frac{1}{1 \pm \frac{m^*\alpha}{\hbar^2 k}}\bigg).
\end{align}
We can write the relaxation times as a single \textit{matrix} in the 2D manifold
\begin{equation}\label{Rashbataupm}
\tau_\pm = \frac{\hbar^3} {n_i m^* U_0^2} \, \bigg(1 \mp \frac{m^*\alpha}{\hbar^2 k} \bigg)\ \rightarrow \ \tau = \tau_0 \, \bigg(\openone - \frac{m^*\alpha}{\hbar^2 k} \, \sigma_z\bigg),
\end{equation}
where $\tau_0 = \hbar^3/(n_i m^* U_0^2)$. We keep only terms up to first order in $\alpha$ and we define $\varepsilon^0_{{\bm k}} = \hbar^2 k^2/2m^*$. The non-equilibrium spin polarisation stems from
\begin{equation}
\arraycolsep 0.3ex
\begin{array}{rl}
\displaystyle n^{(-1)}_{E\bm k} \approx & \displaystyle \frac{e\alpha m^*{\bm E} \cdot\hat{\bm k} \tau_0}{\hbar^3}\, \sigma_z \, \left[\frac{1}{k} \delta(k - k_F) - \pd{}{k} \delta(k - k_F)\right].
\end{array}
\end{equation}
Here $k_F = (2\pi n)^{1/2}$ with $n$ the electron number density. Without loss of generality consider ${\bm E} \parallel \hat{\bm x}$.
In the eigenstate basis, the spin-$y$ operator is given by $s_y=-\hbar/2(\cos\theta\sigma_z+\sin\theta\sigma_y)$.
Then the expectation value of the $y$-component of the spin is obtained as
\begin{equation}
\arraycolsep 0.3ex
\begin{array}{rl}
\bkt{s_y}=\displaystyle {\rm Tr} \, [s_y n^{(-1)}_{E\bm k}] = & \displaystyle - \frac{e\alpha m^*E_x \tau_0}{2\pi \hbar^2},
\end{array}
\end{equation}
as expected \cite{Inoue_RashbaSHE_Vertex_PRB04,Comment-B}.

\subsubsection{Anomalous driving term and vanishing spin-Hall conductivity}

We compute the spin-Hall conductivity of the system.
The spin-Hall conductivity comes from the off-diagonal part of the density matrix.
First we calculate the contributions from the Berry connection $\bm{\mathcal{R}}_{\bm{k}}$. According to
Eq.~(\ref{Rashba_eignve}), we obtain
\begin{equation}
\bm{\mathcal{R}}_{\bm{k}} = \frac{\hat{\bm{\theta}}}{2 k} \left( \openone +\sigma_{x} \right),
\end{equation}
where $\hat{\bm{\theta}} = (-\sin\theta, \cos\theta)$ is the polar unit vector in reciprocal space. Then intrinsic off-diagonal driving term becomes
\begin{equation}
D_{E\bm{k}} = - \frac{e \bm{E} \cdot \hat{\bm{\theta}}}{2\hbar^{2} k}
\left[ f_0(\varepsilon_{\bm{k}}^+) - f_0(\varepsilon_{\bm{k}}^-) \right] \sigma_{y},
\end{equation}
from which the intrinsic off-diagonal density matrix is obtained as
\begin{equation}
S^{(0)int}_{E\bm{k}} = \frac{e \bm{E} \cdot \hat{\bm{\theta}}}{4\hbar \alpha k^2} \left[ f_0(\varepsilon_{\bm{k}}^+) - f_0(\varepsilon_{\bm{k}}^-) \right] \sigma_{x}.
\end{equation}
Next we calculate the Fermi surface contribution, i.e., the contribution from the anomalous driving term $D'_{E{\bm k}} = - J_{od}[n_{E{\bm k}}^{(-1)}]$.
The matrix elements of the scattering potential are given by $U^{++}_{\bm{k}\bm{k}'}U^{+-}_{\bm{k}'\bm{k}}=(iU_0^2/2)\sin\gamma$ and $U^{+-}_{\bm{k}\bm{k}'}U^{--}_{\bm{k}'\bm{k}}=-(iU_0^2/2)\sin\gamma$.
Here, we have defined $\gamma=\theta'-\theta$. For $n^{(-1)}_{E\bm k}$ we use Eq.~(\ref{Rashbarhopm}), yielding the off-diagonal scattering terms from Eq.~(\ref{PJ}) (see Appendix \ref{Appendix-B} for details)
\begin{align}
\label{J_od-Rashba}
J_{od} [n^{(-1)}_{E\bm k}]^{+-} & = \frac{i e E_x \alpha m^{*} \sin \theta}{\hbar^{3} k} \delta ( k - k_F), \nonumber\\
J_{od} [n^{(-1)}_{E\bm k}]^{-+} & = -\frac{i e E_x \alpha m^{*} \sin \theta}{\hbar^{3} k} \delta ( k - k_F).
\end{align}
Here, we have expanded the delta functions in $J_{od} (n_{E\bm k})$ and retained the terms up to linear order in $\alpha$ as $\delta(\varepsilon_{\bm{k}}^+-\varepsilon_{\bm{k}'}^-)\approx \delta(\varepsilon_{\bm{k}}^0-\varepsilon_{\bm{k}'}^0)+\alpha(k+k')\partial  \delta(\varepsilon_{\bm{k}}^0-\varepsilon_{\bm{k}'}^0)/\partial \varepsilon_{\bm{k}}^0$.
Then we obtain the the intrinsic off-diagonal density matrix as
\begin{equation}
S^{(0)ext}_{E\bm{k}} = -\frac{e E_x m^*\sin\theta}{2\hbar^2 k^2} \delta ( k - k_F)\, \sigma_{x}.
\end{equation}

The spin current operator (in the $y$ direction) is given by $j_s^z=\frac{1}{2}(s_z v_y+v_y s_z)=(\hbar^2 k_y/m)s_z$.
In the eigenstate basis, we have $s_z=(\hbar/2)\sigma_x$.
With the use of the expression for the off-diagonal part of the density matrix (\ref{rhood}), the spin-Hall current is given by $j_y^z=\mathrm{Tr}[j_s^z S^{(0)}_{E\bm{k}}]$, where $S^{(0)}_{E\bm{k}}=S^{(0)int}_{E\bm{k}}+S^{(0)ext}_{E\bm{k}}$.
We find that the spin-Hall current contributions from the intrinsic driving term $D_{E{\bm k}}$ [$j_{y}^{z}(D_{E{\bm k}})=eE_x/(8\pi)$] and the extrinsic (anomalous) driving term $D'_{E{\bm k}}$ [$j_{y}^{z}(D'_{E{\bm k}})=-eE_x/(8\pi)$] cancel out, yielding a zero total spin-Hall current
\begin{equation}
\arraycolsep 0.3ex
\begin{array}{rl}
\displaystyle j_{y}^{z} = 0,
\end{array}
\end{equation}
as expected \cite{Inoue_RashbaSHE_Vertex_PRB04}.

\subsection{Anomalous Hall effect in paramagnetic semiconductors with linear Rashba spin-orbit coupling}

We now consider the case in which a magnetization $M$ exists in the system, pointing along the $\hat{\bm z}$-axis.
The Hamiltonian of the system reads
\begin{equation}
H_{0{\bm k}} = \hbar^2 k^2/2m^* + \alpha (\sigma_x k_y - \sigma_y k_x)+M\sigma_z.
\end{equation}
The energies are $\varepsilon^\pm_{\bm k} = \hbar^2k^2/2m^* \pm \lambda_k$ with $\lambda_k=\sqrt{\alpha^2 k^2 + M^2}$, where $M$ for simplicity has units of energy. We define the two Fermi wave vectors $k_{F\pm}$ by setting $\varepsilon^\pm_{\bm k} = \varepsilon_F$. As $M \rightarrow 0$ they both tend to a common value referred to as $k_F$. Below we will assume $\alpha k_F, M \ll \varepsilon_F$, but no assumptions are made about the relative size of $\alpha k_F$ and $M$. The eigenvectors are
\begin{equation}
\arraycolsep 0.3ex
\begin{array}{rl}
\displaystyle \ket{u^+_{\bm{k}}} = & \displaystyle \frac{1}{\sqrt{2\lambda_k}} \, \begin{bmatrix}
\frac{\alpha k}{\sqrt{\lambda_k - M}} \, e^{-i\theta} \cr
- i \sqrt{\lambda_k - M}
\end{bmatrix}, \\ [3ex]

\displaystyle \ket{u^-_{\bm{k}}} = & \displaystyle \frac{1}{\sqrt{2\lambda_k}} \, \begin{bmatrix}
\frac{\alpha k}{\sqrt{\lambda_k + M}} \, e^{-i\theta} \cr
i \sqrt{\lambda_k + M}
\end{bmatrix}.
\end{array}
\end{equation}
The Berry connection is
\begin{equation}
\bm{\mathcal{R}}_{\bm{k}} = \frac{\hat{\bm{\theta}}}{2 k} \left( \openone +\frac{M}{\lambda_k}\sigma_z+\frac{\alpha k}{\lambda_k}\sigma_{x} \right).
\end{equation}
The intrinsic off-diagonal driving term becomes
\begin{equation}
D_{E\bm{k}} = - \frac{e \bm{E} \cdot \hat{\bm{\theta}}}{2\hbar k}\frac{\alpha k}{\lambda_k}
\left[ f_0(\varepsilon_{\bm{k}}^+) - f_0(\varepsilon_{\bm{k}}^-) \right] \sigma_{y},
\end{equation}
from which the intrinsic off-diagonal density matrix is obtained as
\begin{equation}
S^{(0)int}_{E\bm{k}} = \frac{e \bm{E} \cdot \hat{\bm{\theta}}\alpha}{4\lambda^2_k} \left[ f_0(\varepsilon_{\bm{k}}^+) - f_0(\varepsilon_{\bm{k}}^-) \right] \sigma_{x}.
\end{equation}
Without loss of generality consider ${\bm E} \parallel \hat{\bm x}$.
Then the intrinsic anomalous Hall conductivity from the Berry phase term is calculated to be
\begin{align}
\sigma^0_{xy} &= \mathrm{Tr}[(-e)v_y S^{(0)int}_{E\bm{k}}]/E_x \nonumber\\
&= -\frac{e^2M}{2h}\bigg(\frac{1}{\sqrt{\alpha^2k^2_{F+}+M^2}}-\frac{1}{\sqrt{\alpha^2k^2_{F-}+M^2}}\bigg).
\end{align}
In the regime $M \ll \alpha k_F$, we get
\begin{equation}
\sigma^0_{xy} = \frac{e^2M}{2h}\frac{2\alpha^2m^*}{\hbar^2\lambda^2_{k_F}},
\end{equation}
where $\lambda_{k_F} \equiv \lambda_k (k = k_F)$, $v_y=\partial H_{0\bm{k}}/\partial k_y$ is the velocity operator in the eigenstate basis, and $k_{F+} - k_{F-} \approx 2\alpha m^*/\hbar^2$.

Next we compute the contribution to the anomalous Hall conductivity from the anomalous driving term $D'_{E{\bm k}} = - J_{od}[n_{E{\bm k}}^{(-1)}]$.
We consider a short-range (on-site) disorder of the form $U(\bm{r})=U_0\sum_i\delta(\bm{r}-\bm{r}_i)$, and assume that the correlation function satisfies $\langle U(\bm{r})U(\bm{r}')\rangle=n_{i}U_0^2\,\delta(\bm{r}-\bm{r}')$ with $n_{i}$ the impurity density.
The diagonal part of the density matrix takes the form
\begin{equation}\label{rhodRashbaAHE}
n^{(-1),nn}_{E{\bm k}} = - \frac{e{\bm E}\cdot\hat{\bm k} \tau_n}{\hbar} \pd{\varepsilon^n_{\bm k}}{k} \, \delta(\varepsilon^n_{\bm k} - \varepsilon_F).
\end{equation}
Once more we write the relaxation times as a matrix
\begin{equation}\label{RashbaAHEtaupm}
\tau = \tau_0 \, \left(\openone - \frac{m^*\alpha^2}{\hbar^2 \sqrt{\alpha^2 k^2 + M^2}} \, \sigma_z\right),
\end{equation}
where $\tau_0 = \hbar^3/(n_i m^* U_0^2)$. This tends to the result for the non-magnetic case when $M = 0$, and to the correct, $M$-independent result when $\alpha = 0$. Intuitively, for $\alpha = 0$ and $M \ne 0$ the eigenstates are pure spin up and down, and the scalar potential cannot give interband scattering, which would be equivalent to spin-flip. In that limit $U^{\pm\pm}$ tends to a constant and $U^{\pm\mp} = 0$. The anomalous driving term can be evaluated in the same way as Eq. (\ref{J_od-Rashba}) by expanding the delta functions and retaining the terms up to linear order in $\lambda_k$.
The final form reads
\begin{equation}
J_{od} [n^{(-1)}_{E\bm k}]^{+-} = \left(\frac{eE_x \alpha Mm^*\cos\theta}{\lambda_k k\hbar^3} + \frac{ieE_x \alpha m^*\sin\theta}{ k\hbar^3}\right) \, \delta(k - k_F),
\end{equation}
from which the extrinsic off-diagonal density matrix is obtained as
\begin{equation}
S^{(0)ext}_{E\bm{k}} = -\delta(k-k_F)\left[\frac{eE_x \alpha m^*M\cos\theta}{2\lambda^2_k k\hbar^2}\sigma_y+\frac{eE_x \alpha m^*\sin\theta}{2\lambda_k k\hbar^2}\sigma_x\right],
\end{equation}
This extrinsic off-diagonal density matrix contributes to the anomalous Hall conductivity as
 \begin{equation}
 \arraycolsep 0.3ex
\begin{array}{rl}
\displaystyle \sigma'_{xy} =\mathrm{Tr}[(-e)v_y S^{(0)ext}_{E\bm{k}}]/E_x= -\frac{e^2}{2h}\frac{2\alpha^2m^*M}{\hbar^2\lambda^2_{k_F}}.
\end{array}
 \end{equation}
In the regime $M \ll \alpha k_F$, we have vanishing total anomalous Hall conductivity due to the cancellation of the intrinsic and extrinsic contributions
\begin{equation}
\sigma^0_{xy} + \sigma'_{xy} = 0,
\end{equation}
as expected \cite{AHE_vertex_PRL_2006, Nunner2007}.

\section{Conclusions}

We have described a general way of evaluating the linear response of a crystal to an electric field, accounting for both the intra-band and inter-band response. The intra-band response is captured by the band-diagonal part of the density matrix, in the Born approximation is inversely proportional to the strength of the disorder potential to leading order, and is responsible for properties such as the longitudinal conductivity and current-induced spin polarizations in spin-orbit coupled systems. Inter-band coherence makes a sizable contribution to the linear response of crystals. It is captured by the band off-diagonal part of the density matrix, it has strong intrinsic as well as extrinsic contributions, and its leading contribution is of zeroth order in the disorder potential. It is responsible for properties such as the spin-Hall effect, the anomalous Hall effect, the minimum conductivity of massless Dirac fermions. In particular, our theory can capture correctly chiral-anomaly-induced transport phenomena such as the negative magnetoresistance of Weyl semimetals \cite{Sekine2017}. The interplay of the diagonal and off-diagonal parts of the density matrix is important in determining the linear response of ferromagnets to an electric field, including contributions such as spin-orbit torques. The method described in this work can be generalized to include extrinsic spin-dependent scattering effects \cite{TseSDS_SJ_PRL06, Culcer2010} and these will be the topic of a future publication.

\acknowledgements

This research was supported by the Australian Research Council Centre of Excellence in Future Low-Energy Electronics Technologies (project number CE170100039) and funded by the Australian Government. Work at Austin was supported by the Department of Energy, Office of Basic Energy Sciences under Contract No.~DE-FG02-ER45958 and by the Welch foundation under Grant No.~TBF1473. A.S. is supported by the JSPS Overseas Research Fellowship. We thank Hong Liu and Weizhe Liu for double-checking our results, as well as Martin Gradhand, Nikitas Gidopoulos, Stuart Clark, Jared Cole, Hua Chen and Haizhou Lu for enlightening discussions.

\appendix

\begin{widetext}

\section{Complete expressions for the scattering term \label{Appendix-A}}
Here we give the complete expressions for the scattering term $J_{\bm{k}}(\langle\rho\rangle)$ [Eq.~(\ref{J-general-form})].
The complete expression for the energy-conserving $\delta$-function scattering terms is
\begin{equation}
\arraycolsep 0.3ex
\begin{array}{rl}
\displaystyle J_\delta(\bkt{\rho})^{mm'''}_{{\bm k}} = & \displaystyle \frac{\pi}{\hbar} \sum_{m'm''{\bm k}'} \left\{ \bkt{U^{mm'}_{{\bm k}{\bm k}'} U^{m'm''}_{{\bm k}'{\bm k}}} \bkt{\rho}^{m''m'''}_{{\bm k}} \delta(\varepsilon^{m'}_{{\bm k}'} - \varepsilon^{m''}_{{\bm k}}) + \bkt{U^{m'm''}_{{\bm k}{\bm k}'} U^{m''m'''}_{{\bm k}'{\bm k}}} \bkt{\rho}^{mm'}_{\bm k} \delta(\varepsilon^{m'}_{\bm k} - \varepsilon^{m''}_{{\bm k}'})\right. \\ [3ex]
&\left. - \displaystyle \bkt{U^{mm'}_{{\bm k}{\bm k}'} U^{m''m'''}_{{\bm k}'{\bm k}}} \bkt{\rho}^{m'm''}_{{\bm k}'} \delta(\varepsilon^{m''}_{{\bm k}'} - \varepsilon^{m'''}_{{\bm k}}) - \bkt{U^{mm'}_{{\bm k}{\bm k}'} U^{m''m'''}_{{\bm k}'{\bm k}}} \bkt{\rho}^{m'm''}_{{\bm k}'} \delta(\varepsilon^m_{\bm k} - \varepsilon^{m'}_{{\bm k}'}) \right\},
\end{array}
\end{equation}
where we have used $J_\delta = J_d + J_{od}$, as in Eq.\ (\ref{PJ}). The principal part terms take the form
\begin{equation}
\arraycolsep 0.3ex
\begin{array}{rl}
\displaystyle J_{pp}(\bkt{\rho})^{mm'''}_{{\bm k}} = & \displaystyle \frac{1}{i\hbar} \mathcal{P} \sum_{m'm''{\bm k}'}  \left\{ \bkt{U^{mm'}_{{\bm k}{\bm k}'} U^{m'm''}_{{\bm k}'{\bm k}}} \bkt{\rho}^{m''m'''}_{{\bm k}} \frac{1}{\varepsilon^{m'}_{{\bm k}'} - \varepsilon^{m''}_{{\bm k}}} + \bkt{U^{m'm''}_{{\bm k}{\bm k}'} U^{m''m'''}_{{\bm k}'{\bm k}}} \bkt{\rho}^{mm'}_{\bm k} \frac{1}{\varepsilon^{m'}_{\bm k} - \varepsilon^{m''}_{{\bm k}'}}\right. \\ [3ex]

&\left. -  \displaystyle \bkt{U^{mm'}_{{\bm k}{\bm k}'} U^{m''m'''}_{{\bm k}'{\bm k}}} \bkt{\rho}^{m'm''}_{{\bm k}'} \frac{1}{\varepsilon^{m''}_{{\bm k}'} - \varepsilon^{m'''}_{{\bm k}}} - \bkt{U^{mm'}_{{\bm k}{\bm k}'} U^{m''m'''}_{{\bm k}'{\bm k}}} \bkt{\rho}^{m'm''}_{{\bm k}'} \frac{1}{\varepsilon^m_{\bm k} - \varepsilon^{m'}_{{\bm k}'}} \right\}.
\end{array}
\end{equation}

To get the corresponding expressions for the band off-diagonal part of the density matrix, $S^{mm'}_{\bm k}$, one simply replaces $\bkt{\rho} \rightarrow S_{\bm k}$ in the above.

\section{Detailed derivation of the anomalous driving term in the Rashba model \label{Appendix-B}}
Here we show a detailed derivation of Eq.~(\ref{J_od-Rashba}), the anomalous driving term $D'_{E{\bm k}} = - J_{od}[n_{E{\bm k}}^{(-1)}]$ in the Rashba model.
The spin-Hall conductivity comes from the off-diagonal part of the density matrix. The anomalous driving term arises from the following scattering term [Eq.~(\ref{PJ})]
\begin{equation}
\arraycolsep 0.3ex
\begin{array}{rl}
\displaystyle J_{od} [n^{(-1)}_{E\bm{k}}]^{+-} = & \displaystyle \frac{\pi}{\hbar} \sum_{{\bm k}'} \bkt{U^{++}_{{\bm k}{\bm k}'} U^{+-}_{{\bm k}'{\bm k}}} \left\{ \left[n^{(-1),++}_{E\bm{k}} - n^{(-1),++}_{E\bm{k}'}\right] \delta(\varepsilon^+_{\bm k} - \varepsilon^{+}_{{\bm k}'}) + \left[n^{(-1),--}_{E\bm{k}} - n^{(-1),++}_{E\bm{k}'}\right] \delta(\varepsilon^{-}_{{\bm k}} - \varepsilon^{+}_{{\bm k}'})\right\} \\ [3ex]
& + \displaystyle \frac{\pi}{\hbar} \sum_{{\bm k}'} \bkt{U^{+-}_{{\bm k}{\bm k}'} U^{--}_{{\bm k}'{\bm k}}} \left\{ \left[n^{(-1),++}_{E\bm{k}} - n^{(-1),--}_{E\bm{k}'}\right] \delta(\varepsilon^+_{\bm k} - \varepsilon^{-}_{{\bm k}'}) + \left[n^{(-1),--}_{E\bm{k}} - n^{(-1),--}_{E\bm{k}'}\right] \delta(\varepsilon^{-}_{{\bm k}} - \varepsilon^{-}_{{\bm k}'})\right\}.
\end{array}
\end{equation}
We use Eq.\ (\ref{Rashbarhopm}) for $n^{(-1),\pm\pm}_{E\bm{k}}$, and the matrix elements of the scattering potential are
\begin{align}
U^{++}_{{\bm k}{\bm k}'}U^{+-}_{{\bm k}'{\bm k}} =\frac{iU^2}{2} \, \sin\gamma,\ \ \ \ \ \ \ \ 
U^{+-}_{{\bm k}{\bm k}'}U^{--}_{{\bm k}'{\bm k}} = - \frac{iU^2}{2} \, \sin\gamma,
\end{align}
where $\gamma = \theta' - \theta$, and we have used $U^{\pm\pm}_{{\bm k}{\bm k}'} = U/2(e^{-i\gamma} + 1)$ and $U^{\pm\mp}_{{\bm k}{\bm k}'} = U/2(e^{-i\gamma} - 1)$. Only the terms with $n^{(-1),\pm\pm}_{E\bm{k}'}$ survive the angular integration:
\begin{equation}
\arraycolsep 0.3ex
\begin{array}{rl}
\displaystyle J_{od} [n^{(-1)}_{E\bm{k}}]^{+-} = \frac{i\pi n_{i} U^2}{2\hbar} \sum_{{\bm k}'} \sin\gamma \left\{ n^{(-1),--}_{E\bm{k}'} [\delta(\varepsilon^+_{\bm k} - \varepsilon^{-}_{{\bm k}'}) + \delta(\varepsilon^{-}_{{\bm k}} - \varepsilon^{-}_{{\bm k}'})] - n^{(-1),++}_{E\bm{k}'} [\delta(\varepsilon^+_{\bm k} - \varepsilon^{+}_{{\bm k}'}) + \delta(\varepsilon^{-}_{{\bm k}} - \varepsilon^{+}_{{\bm k}'})] \right\}.
\end{array}
\end{equation}
We substitute explicit forms of $n^{(-1),\pm\pm}_{E\bm{k}'}$ from Eq.\ (\ref{Rashbarhopm}), yielding
\begin{equation}
\arraycolsep 0.3ex
\begin{array}{rl}
\displaystyle J_{od} [n^{(-1)}_{E\bm{k}}]^{+-} = & \displaystyle \frac{i\pi n_{i} U^2}{2\hbar} \frac{e{\bm E}}{\hbar} \cdot \sum_{{\bm k}'} \sin\gamma \left\{ - \tau_- \pd{\varepsilon^-_{{\bm k}'}}{{\bm k}'} \, \delta(\varepsilon^-_{{\bm k}'} - \varepsilon_F) [\delta(\varepsilon^+_{\bm k} - \varepsilon^{-}_{{\bm k}'}) + \delta(\varepsilon^{-}_{{\bm k}} - \varepsilon^{-}_{{\bm k}'})] \right.\\ [3ex]
&\left. + \displaystyle \tau_+ \pd{\varepsilon^+_{{\bm k}'}}{{\bm k}'} \, \delta(\varepsilon^+_{{\bm k}'} - \varepsilon_F) [\delta(\varepsilon^+_{\bm k} - \varepsilon^{+}_{{\bm k}'}) + \delta(\varepsilon^{-}_{{\bm k}} - \varepsilon^{+}_{{\bm k}'})] \right\}.
\end{array}
\end{equation}
We can turn the summations into integrations, and do the integrals over angle first, taking ${\bm E} \parallel \hat{\bm x}$ and noting that
\begin{equation}
\int\frac{d\theta'}{2\pi} \, \cos\theta' \sin\gamma = -\frac{\sin\theta}{2}.
\end{equation}
This gives
\begin{equation}
\arraycolsep 0.3ex
\begin{array}{rl}
\displaystyle J_{od} [n^{(-1)}_{E\bm{k}}]^{+-} = & \displaystyle \frac{i\pi n_{i} U^2}{8\pi \hbar} \frac{eE\sin\theta}{\hbar} \int dk' \, k' \, \left\{ \tau_- \pd{\varepsilon^-_{{\bm k}'}}{k'} \, \delta(\varepsilon^-_{{\bm k}'} - \varepsilon_F) [\delta(\varepsilon^+_{\bm k} - \varepsilon^{-}_{{\bm k}'}) + \delta(\varepsilon^{-}_{{\bm k}} - \varepsilon^{-}_{{\bm k}'})] \right.\\ [3ex]
&\left. - \displaystyle \tau_+ \pd{\varepsilon^+_{{\bm k}'}}{k'} \, \delta(\varepsilon^+_{{\bm k}'} - \varepsilon_F) [\delta(\varepsilon^+_{\bm k} - \varepsilon^{+}_{{\bm k}'}) + \delta(\varepsilon^{-}_{{\bm k}} - \varepsilon^{+}_{{\bm k}'})] \right\}.
\end{array}\label{J_od-appendix}
\end{equation}
Let us evaluate the $\tau_-$ term in Eq.~(\ref{J_od-appendix}):
\begin{equation}
\arraycolsep 0.3ex
\begin{array}{rl}
\displaystyle J_{od}(\tau_-) = & \displaystyle \frac{i\pi n_{i} U^2\tau_0}{8\pi \hbar} \frac{eE\sin\theta}{\hbar} \int dk' \, \frac{\hbar^2k'^{2}}{m^*} \bigg(1 + \frac{m^*\alpha}{\hbar^2 k'} \bigg) \bigg( 1 - \frac{m^*\alpha}{\hbar^2k'} \bigg) \, \delta(\varepsilon^-_{{\bm k}'} - \varepsilon_F) [\delta(\varepsilon^+_{\bm k} - \varepsilon^{-}_{{\bm k}'}) + \delta(\varepsilon^{-}_{{\bm k}} - \varepsilon^{-}_{{\bm k}'})] \\ [3ex]
= & \displaystyle \frac{i eE \hbar^3 \sin\theta}{8 m^{*2}} \int dk' \, k'^{2} \, \delta(\varepsilon^-_{{\bm k}'} - \varepsilon_F) [\delta(\varepsilon^+_{\bm k} - \varepsilon^{-}_{{\bm k}'}) + \delta(\varepsilon^{-}_{{\bm k}} - \varepsilon^{-}_{{\bm k}'})]+\mathcal{O}(\alpha^2).
\end{array}
\end{equation}
We expand the delta functions up to the first order in the spin-orbit coupling strength $\alpha$ as
\begin{equation}
\arraycolsep 0.3ex
\begin{array}{rl}
\displaystyle \delta(\varepsilon^+_{\bm k} - \varepsilon^{-}_{{\bm k}'}) &\approx \delta(\varepsilon^0_{\bm k} - \varepsilon^{0}_{{\bm k}'}) + \alpha(k + k') \, \pd{}{\varepsilon^0_{\bm k}} \delta(\varepsilon^0_{\bm k} - \varepsilon^{0}_{{\bm k}'}), \\ [2ex]

\displaystyle \delta(\varepsilon^-_{\bm k} - \varepsilon^{-}_{{\bm k}'}) &\approx \delta(\varepsilon^0_{\bm k} - \varepsilon^{0}_{{\bm k}'}) - \alpha(k - k') \, \pd{}{\varepsilon^0_{\bm k}} \delta(\varepsilon^0_{\bm k} - \varepsilon^{0}_{{\bm k}'}),\\ [2ex]

\displaystyle \delta(\varepsilon^+_{\bm k} - \varepsilon^{-}_{{\bm k}'}) + \delta(\varepsilon^-_{\bm k} - \varepsilon^{-}_{{\bm k}'}) &\approx 2\delta(\varepsilon^0_{\bm k} - \varepsilon^{0}_{{\bm k}'}) + 2 \alpha k' \, \pd{}{\varepsilon^0_{\bm k}} \delta(\varepsilon^0_{\bm k} - \varepsilon^{0}_{{\bm k}'}), \\ [2ex]

\displaystyle \delta(\varepsilon^{-}_{{\bm k}'} - \varepsilon_F) &\approx \delta(\varepsilon^0_{{\bm k}'} - \varepsilon_F) - \alpha k' \, \pd{}{\varepsilon^0_{{\bm k}'}} \delta(\varepsilon^0_{{\bm k}'} - \varepsilon_F).
\end{array}
\end{equation}
We only need to evaluate the terms of first order in $\alpha$ out of the above expression, which gives
\begin{equation}
\arraycolsep 0.3ex
\begin{array}{rl}
J_{od}(\tau_-) = & \displaystyle \frac{i e E\alpha \hbar^3 \sin\theta}{4 m^{*2}} \int dk' \, k^{'3} \, \bigg\{\delta(\varepsilon^0_{{\bm k}'} - \varepsilon_F) \, \pd{}{\varepsilon^0_{\bm k}} \delta(\varepsilon^0_{\bm k} - \varepsilon^{0}_{{\bm k}'}) - \delta(\varepsilon^0_{\bm k} - \varepsilon^{0}_{{\bm k}'}) \pd{}{\varepsilon^0_{{\bm k}'}} \delta(\varepsilon^0_{{\bm k}'} - \varepsilon_F) \bigg\} \\ [2ex]

= & \displaystyle \frac{i e E \alpha\hbar^3 \sin\theta}{4 m^{*2}} \bigg\{ \pd{}{\varepsilon^0_{\bm k}} \int dk' \, k^{'3} \, \delta(\varepsilon^0_{{\bm k}'} - \varepsilon_F) \, \delta(\varepsilon^0_{\bm k} - \varepsilon^{0}_{{\bm k}'}) \}

- \ \int dk' \, k^{'3} \, \delta(\varepsilon^0_{\bm k} - \varepsilon^{0}_{{\bm k}'}) \pd{}{\varepsilon^0_{{\bm k}'}} \delta(\varepsilon^0_{{\bm k}'} - \varepsilon_F) \bigg\} \\ [2ex]

= & \displaystyle \frac{i e E\alpha \hbar^3 \sin\theta}{4 m^{*2}} \frac{m^{*3}}{\hbar^6} \bigg\{ \frac{1}{k}\pd{}{k} \int dk' \, k' \, \delta(k' - k_F) \, \delta(k' - k) \}

- \ \int dk' \, k^{'2} \, \delta(k' - k) \pd{}{k'} \frac{1}{k'} \delta(k' - k_F) \bigg\} \\ [2ex]

= & \displaystyle \frac{i e E  \alpha m^{*} \sin \theta}{2 \hbar^{3} k} \delta ( k - k_{F} ).
\end{array}
\end{equation}
We can easily see that the $\tau_{+}$ term in Eq.~(\ref{J_od-appendix}) will give the same result as the $\tau_{-}$ term. Thus we arrive at Eq.~(\ref{J_od-Rashba}):
\begin{equation}
J_{od} [n^{(-1)}_{E\bm{k}}]^{+-}= \frac{i e E \alpha m^{*} \sin \theta}{\hbar^{3} k} \delta ( k - k_{F} ),\ \ \ \ \ \ \ \ 
J_{od} [n^{(-1)}_{E\bm{k}}]^{-+}= - \frac{i e E \alpha m^{*} \sin \theta}{\hbar^{3} k} \delta ( k - k_{F} ),
\end{equation}
where we have used the Hermiticity of the scattering term to get $J_{od} [n^{(-1)}_{E\bm{k}}]^{-+}$.

\end{widetext}

 \bibliographystyle{apsrev4-1}

\begin{thebibliography}{121}
\bibitem [{Com({\natexlab{b}})}]{Comment-C}%
  \BibitemOpen
\bibfield  {journal} {  }\href@noop {} {} ({\natexlab{b}}),\ \bibinfo {note}
  {traditionally the Keldysh method starts out with the Schr\"{o}dinger
  equation, which is fully equivalent to the quantum Liouville
  equation.}\BibitemShut {Stop}%
\end{thebibliography}

%

\end{document}